\pdfoutput=1

\documentclass[a4paper,11pt]{article}
\usepackage{amsfonts}
\usepackage{amsmath}
\usepackage{amsthm}
\usepackage{booktabs}
\usepackage{bm}
\usepackage{bbm}
\usepackage{cases}
\usepackage{dsfont}
\usepackage{eucal}
\usepackage{enumerate}
\usepackage{enumitem}
\usepackage{float}
\usepackage[T1]{fontenc}
\usepackage[hmargin = 2cm, vmargin = 3cm]{geometry}
\usepackage{epsfig}
\usepackage{natbib}

\usepackage{hyperref}

\theoremstyle{plain}

\def\T{{ \mathrm{\scriptscriptstyle T} }}

\def\var{\text{var}}
\def\iid{\overset{\text{iid}}{\sim}}
\def\dif{\mathrm{d}}

 \let\oldthebibliography=\thebibliography
 \let\oldendthebibliography=\endthebibliography
 \renewenvironment{thebibliography}[1]{%
   \footnotesize
   \oldthebibliography{#1}%
   \setlength{\parskip}{0mm}%
   \setlength{\itemsep}{0mm}%
 }{\oldendthebibliography}

\newcommand{\balpha}{{\alpha}}

\newcommand{\btheta}{{\theta}}

\newtheorem{example}{Example}

\newtheorem{proposition}{Proposition}
\newtheorem{definition}{Definition}

\date{}
\begin{document} 
\title{\vspace{-1cm}On the Geometry of Bayesian Inference}
\author{Miguel \textsc{de Carvalho}, Garritt \textsc{L.~Page}, and Bradley \textsc{J.~Barney}}
\date{}
\maketitle 

\begin{abstract}\footnotesize
  We provide a geometric interpretation to Bayesian inference that allows us to introduce a natural measure of the level of agreement between priors, likelihoods, and posteriors. The starting point for the construction of our geometry is the simple observation that the marginal likelihood can be regarded as an inner product between the prior and the likelihood. A key concept in our geometry is that of compatibility, a measure which is based on the same construction principles as Pearson correlation, but which can be used to assess how much the prior agrees with the likelihood, to gauge the sensitivity of the posterior to the prior, and to quantify the coherency of the opinions of two experts. Estimators for all the quantities involved in our geometric setup are discussed, which can be directly computed from the posterior simulation output. Some examples are used to illustrate our methods, including data related to  on-the-job drug usage, midge wing length, and prostate cancer.  \\

\noindent \textsc{keywords}: Bayesian inference; Geometry; Harmonic mean estimator;  Hilbert spaces; Marginal likelihood.
\end{abstract}
\let\thefootnote\relax\footnotetext{\scriptsize Miguel de Carvalho is Assistant Professor of Statistics, School of Mathematics, The University of Edinburgh, UK \textit{(e-mail: \href{mailto:miguel.decarvalho@ed.ac.uk}{miguel.decarvalho@ed.ac.uk})}. 
Garritt L.~Page is Assistant Professor of Statistics, Department of Statistics, Brigham Young University, Provo, Utah \textit{(e-mail: \href{mailto:page@stat.byu.edu}{page@stat.byu.edu})}. Bradley J.~Barney is Visiting Assistant Professor of Statistics, Department of Statistics, Brigham Young University, Provo, Utah \textit{(e-mail: \href{mailto:barney@stat.byu.edu}{barney@stat.byu.edu})}. We thank the Editor, the Associate Editor, and a Reviewer for insighful comments on a previous version of the paper. We extend our thanks to J.~Quinlan for research assistantship and discussions, and to V.~I.~de Carvalho, A.~C.~Davison, D.~Henao, W.~O.~Johnson, A.~Turkman, and F.~Turkman for constructive comments. The research was partially supported by Fondecyt 11121186 and 11121131 and by FCT (Funda\c c\~ao para a Ci\^encia e a Tecnologia) through UID/MAT/00006/2013.}
\section{Introduction}\label{introduction}
Assessing the influence that prior distributions and/or likelihoods have on posterior inference has been a topic of research for some time. One commonly used ad-hoc method suggests fitting a Bayes model using a few competing priors, then visually (or numerically) assessing changes in the posterior as a whole or using some pre-specified posterior summary. More rigorous approaches have also been developed. \cite{Levine:1991} developed a framework to assess sensitivity of posterior inference to sampling distribution (likelihood) and the priors. \cite{B90} introduced the concept of  Bayesian robustness which includes perturbation models (see also \citealt{BB86}). More recently, \cite{EJ11} have compared information available in two competing priors. Related to this work,  \cite{GAL08} advocates the use of so-called weakly informative priors that purposely incorporate less information than available as a means of regularizing. Work has also been dedicated to the so-called prior--data conflict \citep{EM06, WalterAugustin:2009, AE16}. Such conflict can be of interest in a wealth of situations, such as for evaluating how much prior and likelihood information are at odds at the node level in a hierarchical model \citep[see][and references therein]{ScheelGreenRougier:2011}. Regarding sensitivity of the posterior distribution to prior specifications, \cite{LopesTobias:2011} provide a fairly accessible overview.

We argue that a geometric representation of the prior, likelihood, and posterior distribution encourages understanding of their interplay. Considering Bayes methodologies from a geometric perspective is not new, but none of the existing geometric perspectives has been designed with the goal of providing a summary on the agreement or impact that each component of Bayes theorem has on inference and predictions. \cite{A71} used a geometric perspective to build intuition behind each component of Bayes theorem, \cite{ShortleMendel:1996} used a geometric approach to draw conditional distributions in arbitrary coordinate systems, and \cite{AD10} argued that conjugate priors of posterior distributions belong to the same geometry giving an appealing interpretation of hyperparameters. \cite{ZhuIbrahimTang:2011} defined a manifold on which a Bayesian perturbation analysis can be carried out by perturbing data, prior and likelihood simultaneously, and \cite{kurtek2015} provide an elegant geometric construction which allows for Bayesian sensitivity analysis based on the so-called $\epsilon$-compatibility class and on comparison of posterior inferences using the Fisher--Rao metric. 

In this paper, we develop a geometric setup along with a set of metrics that can be used to provide an informative preliminary `snap-shot' regarding comparisons between prior and likelihood (to assess the level of agreement between prior and data), prior and posterior (to determine the influence that prior has on inference), and prior versus prior (to compare `informativeness'---i.e., a density's peakedness---{and/or congruence} of two competing priors). To this end, we treat each component of Bayes theorem as an element of a geometry formally constructed using concepts from Hilbert spaces and tools from abstract geometry. Because of this, it is possible to calculate norms, inner products, and angles between vectors. Not only do each of these numeric summaries have intuitively appealing individual interpretations, but they may also be combined to construct a unitless measure of compatibility, which can be used to assess how much the prior agrees with the likelihood, to gauge the sensitivity of the posterior to the prior, and to quantify the coherency of the opinions of two experts. Estimating our measures of level of agreement is straightforward and can actually be carried out within an MCMC algorithm. An important advantage of our setting is that it offers a direct link to Bayes theorem, and a unified treatment that can be used to assess the level of agreement between priors, likelihoods, and posteriors---or functionals of these. To streamline the illustration of ideas, concepts, and methods we reference the following example \citep[][pp.~26--27]{CAL11} throughout the article. \\ \ \\ \vspace{-.6cm}

\noindent {\textsc{On-the-job drug usage toy example}} \\
Suppose interest lies in estimating the proportion $\theta \in [0,1]$ of US transportation industry workers that use drugs on the job. {Suppose} $n = 10$ workers were selected and tested with the 2nd and 7th testing positive. {Let ${y} = (Y_1, \dots, Y_{n})$ with $Y_i = 1$ denoting that the $i$th worker tested positive and $Y_i = 0$ otherwise. Let $Y_i \mid \theta \overset{\text{iid}}{\sim} \text{Bern}(\theta)$}, for $i = 1, \ldots, n$, and $\theta \sim \text{Beta}(a,b)$, for $a, b > 0$. Then, $\theta \mid {{y}} \sim \text{Beta}(a^{\star}, b^{\star})$ with $a^{\star} = n_1 + a$ and $b^{\star}= n - n_1 + b$, where $n_1 = \sum_{i = 1}^n Y_i$. \\ 

Some natural questions our treatment of Bayes theorem will answer are: How compatible is the likelihood with this prior choice?  How similar are the posterior and prior distributions? How does the choice of $\text{Beta}(a,b)$ compare to other possible prior distributions? While the drug usage example provides a recurring backdrop that we consistently call upon, additional examples are used throughout the paper to illustrate our methods.  

In Section~\ref{Geometry} we introduce the geometric framework in which we work and provide definitions and interpretations along with examples. Section~\ref{extensions} considers extensions of the proposed setup, Section~\ref{estimation} contains computational details, and Section~\ref{Reg} provides a regression example illustrating utility of our metric. Section~\ref{discussion} conveys some concluding remarks. Proofs are given in the supplementary materials. 

\section{Bayes geometry}\label{Geometry}
\subsection{A geometric view of Bayes theorem}
Suppose the inference of interest is over a parameter ${\theta}$ which takes values on $\Theta \subseteq \mathbb{R}^p$. We consider the space of square integrable functions $L_2(\Theta)$, and use the geometry of the Hilbert space $\mathcal{H} = (L_2(\Theta),\langle \cdot, \cdot \rangle)$, with inner-product 
\begin{align} \label{innerproduct}
  \langle g, h \rangle = \int_{\Theta} g (\btheta) h (\btheta)\, \dif \btheta, \quad g,h \in L_2(\Theta).
\end{align}
The fact that $\mathcal{H}$ {is a} Hilbert space is often known in mathematical parlance as the Riesz--Fischer theorem;  for a proof see \citet[][p.~411]{C01}. Borrowing geometric terminology from linear spaces, we refer to the elements of $L_2(\Theta)$ as vectors, and assess their `magnitudes' through the use of the norm induced by the inner product in \eqref{innerproduct}, i.e., $\|\cdot\| = (\langle \cdot, \cdot\rangle)^{1/2}$. 

The starting point for constructing our geometry is the observation that Bayes theorem can be written using the inner-product in \eqref{innerproduct} as follows
\begin{equation}\label{Bayes}
  p(\btheta \mid {y})  = 
  \frac{ \pi(\btheta) f({y} \mid \btheta) }{\int_{\Theta} \pi(\btheta) 
    f({{y}}\mid  \btheta)\, \dif \btheta}
  = \frac{\pi(\btheta) \ell(\btheta) }{\langle \pi, \ell \rangle},
\end{equation}
where $\ell(\btheta) = f({y} \mid \btheta)$ denotes the likelihood, $\pi(\btheta)$ is a prior {density},  $p(\btheta \mid {y})$ is the posterior density and $\langle \pi, \ell \rangle = \int_{\Theta} \ f({y} \mid \btheta) \pi(\btheta) \,\dif \btheta$ is the marginal likelihood or integrated likelihood. The inner product in \eqref{innerproduct} naturally leads to considering $\pi$ and $\ell$ that are in $L_2(\Theta)$, which is compatible with a wealth of parametric models and proper priors. By considering $p$, $\pi$, and $\ell$ as vectors with different magnitudes and directions, Bayes theorem simply indicates how one might recast the prior vector {so as} to obtain the posterior vector. The likelihood vector is used to enlarge/reduce the magnitude and suitably tilt the direction of the prior vector in a sense that will be made precise below. 

The marginal likelihood $\langle \pi, \ell \rangle $ is simply the inner product between the likelihood and the prior, and hence can be understood as a  measure of agreement between the prior and the likelihood. To make this more concrete, define the \emph{angle measure} between the prior and the likelihood as  
\begin{equation}\label{angle}  
\pi \angle \hspace{.05cm}\ell = \arccos \frac{\langle \pi, \ell \rangle}{\|\pi\| \|\ell\|}.
\end{equation}
Since $\pi$ and $\ell$ are nonnegative, the angle between the prior and the likelihood can only be acute or right, i.e., $\pi \angle \hspace{.05cm}\ell \in [0,90^{\circ}]$.  The closer $\pi \angle \hspace{.05cm}\ell$ is to $0^{\circ}$, the greater the agreement between the prior and the likelihood. Conversely, the closer $\pi \angle \hspace{.05cm}\ell$ is to $90^{\circ}$, the greater the disagreement between prior and likelihood. In the pathological case where $\pi \angle \hspace{.05cm}\ell = 90^{\circ}$ (which requires the prior and the likelihood to have all of their mass on disjoint sets), we say that the prior is orthogonal to the likelihood. Bayes theorem is incompatible with a prior being orthogonal to the likelihood as $\pi \angle \hspace{.05cm}\ell =90^{\circ}$ indicates that $\langle \pi, \ell \rangle=0$, thus leading to a division by zero in \eqref{Bayes}. Similar to the correlation coefficient for random variables in $L_2(\Omega, \mathbb{B}_{\Omega}, P)$---with  $\mathbb{B}_{\Omega}$ denoting the Borel sigma-algebra over the sample space $\Omega$---, our target object of interest is given by a standardized inner product 
\begin{equation}\label{kappa.pi}
  \kappa_{\pi,\ell} = \frac{\langle \pi, \ell \rangle}{\|\pi\| \|\ell\|}.
\end{equation}
The quantity $\kappa_{\pi,\ell}$ quantifies how much an expert's opinion agrees with the data, thus providing a natural measure of the level of agreement between prior and data. 



Before exploring \eqref{kappa.pi} more fully by providing interpretations and properties we concretely define how the term `geometry' will be used throughout the paper. The following definition of abstract geometry can be found in \citet[][p.~17]{MP91}.
\begin{definition}[Abstract geometry]\label{absgeom}
  An abstract geometry $\mathcal{A}$ consists of a pair $\{\mathcal{P}, \mathcal{L}\}$, where the elements of set $\mathcal{P}$ are designed as points, and the elements of the collection $\mathcal{L}$ are designed as lines, such that:
  \begin{enumerate}
  \item For every two points $A, B \in \mathcal{P}$, there is a line $l \in \mathcal{L}$.
  \item Every line has at least two points.
  \end{enumerate}
\end{definition}
Our abstract geometry of interest is $\mathcal{A} = \{\mathcal{P}, \mathcal{L}\}$, where $\mathcal{P} = L_2(\Theta)$ and the set of all lines is 
\begin{equation}
  \label{lines}
  \mathcal{L} = \{g + k h: g, h \in L_2(\Theta), k \in \mathbb{R}\}.
\end{equation}
Hence, in our setting points can be, for example, prior densities, posterior densities, or likelihoods, as long as they are in $L_2(\Theta)$. Lines are elements of $\mathcal{L}$, {as defined in \eqref{lines}}, so that for example if $g$ and $h$ are densities, line segments in our geometry consist of all possible mixture distributions which can be obtained from $g$ and $h$, i.e.,
\begin{equation}\label{lambda}
  \{\lambda g + (1-\lambda) h: \lambda \in [0,1]\}.
\end{equation}
{A related interpretation of two-component mixtures as straight lines can be found in \citet[][p.~82]{marriott2002}.} 

Vectors in $\mathcal{A} = \{\mathcal{P}, \mathcal{L}\}$ are defined through the difference of elements in $\mathcal{P} = L_2(\Theta)$. For example, let $g \in L_2(\Theta)$ and let $0 \in L_2(\Theta)$. Then $g = g - 0 \in L_2(\Theta)$, and hence $g$ can be regarded both as a point {and} as a vector. If $g, h \in L_2(\Theta)$ are vectors then we say that $g$ and $h$ are collinear if there exists $k \in \mathbb{R}$, such that $g(\btheta) = k h(\btheta)$. Put differently, we say $g$ and $h$ are collinear if $g(\btheta) \propto h(\btheta)$, for all $\btheta \in \Theta$. 

For any two points in the geometry under consideration, we define their compatibility as a standardized inner product (with \eqref{kappa.pi} being a particular case).
\begin{definition}[Compatibility]\label{Comp.def}
  The compatibility between points in the geometry under consideration is defined as
\begin{equation}\label{compatibility}
  \kappa_{g, h}=\frac{\langle g, h \rangle}{\|g\| \|h\|}, \quad g, h \in L_2(\Theta).
\end{equation}
\end{definition}

The concept of compatibility in Definition~\ref{Comp.def} is based on the same construction principles as the Pearson correlation coefficient, which would be based however on the inner product
\begin{equation}\label{cov}
  \langle X, Y \rangle = \int_{\Omega} X Y \, \dif P, \quad X, Y \in L_2(\Omega, \mathbb{B}_\Omega, P),
\end{equation}
instead of the inner product in \eqref{innerproduct}. However, compatibility is defined for priors, posteriors, and likelihoods in $L_2(\Theta)$ equipped with the inner product \eqref{innerproduct}, whereas Pearson correlation works with random variables in $L_2(\Omega, \mathbb{B}_\Omega, P)$ equipped with the inner product \eqref{cov}. Our concept of compatibility can be used to evaluate how much the prior agrees with the likelihood, to measure the sensitivity of the posterior to the prior, and to quantify the level of agreement of elicited priors. As an illustration consider the following example.

\begin{example}\label{simple.example}\normalfont
Consider the following densities $\pi_0(\theta)=I_{(0,1)}(\theta)$, $\pi_1(\theta)=1 / 2 I_{(0,2)}(\theta)$, $\pi_2(\theta)=I_{(1,2)}(\theta)$, and $\pi_3(\theta)=1 / 2 I_{(1,3)}(\theta)$. Note that $\|\pi_0\|=\|\pi_2\|=1$, $\|\pi_1\|=\|\pi_3\|=\sqrt{2}/2$, and; further,  $\kappa_{\pi_0,\pi_1}=\kappa_{\pi_2,\pi_3}=\sqrt{2}/2$, thus implying that $\pi_0 \angle \pi_1 = \pi_2 \angle \pi_3 = 45^{\circ}$. Also, $\kappa_{\pi_0, \pi_2}=0$ and hence $\pi_0 \perp \pi_2$.
\end{example}

As can be observed in Example~\ref{simple.example}, $(\pi_a \angle \pi_b)/90^{\circ}$ is a natural measure of distinctiveness of two densities. In addition, Example~\ref{simple.example} shows us how different distributions can be  associated to the same norm and angle. Hence, as expected, any Cartesian representation $(x,y) \mapsto (\|\cdot\|\cos(\cdot \angle \cdot), \|\cdot\|\sin(\cdot \angle\cdot))$, will only allow us to represent some features of the corresponding distributions, but will not allow us to identify the distributions themselves. 

\begin{figure}[h] \centering
  \begin{minipage}[c]{0.45\textwidth}   \centering
    \includegraphics[height = 6.3cm]{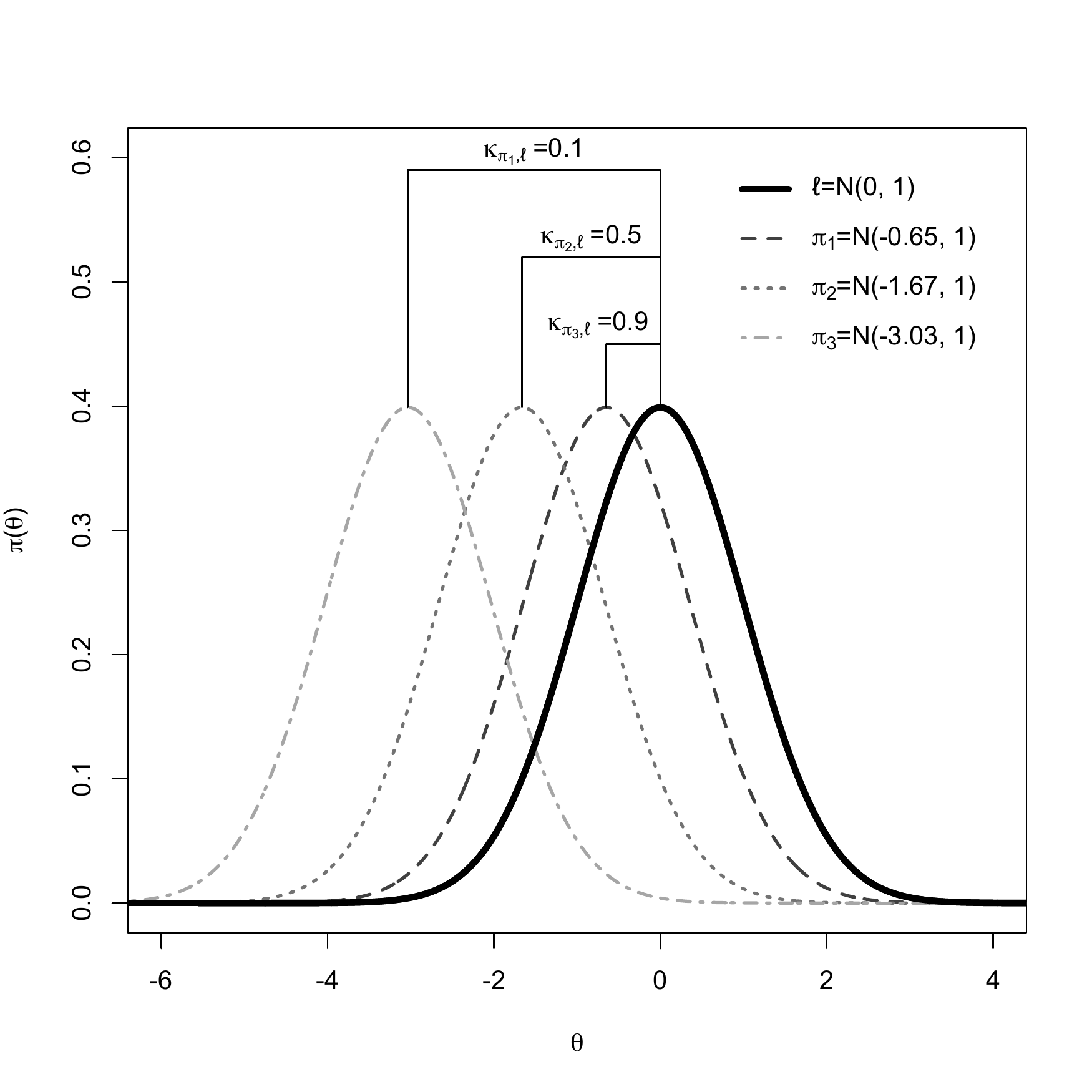} 
  \end{minipage} \hspace{0.7cm}
  \begin{minipage}[c]{0.45\textwidth}   \centering
    \includegraphics[height = 6.3cm]{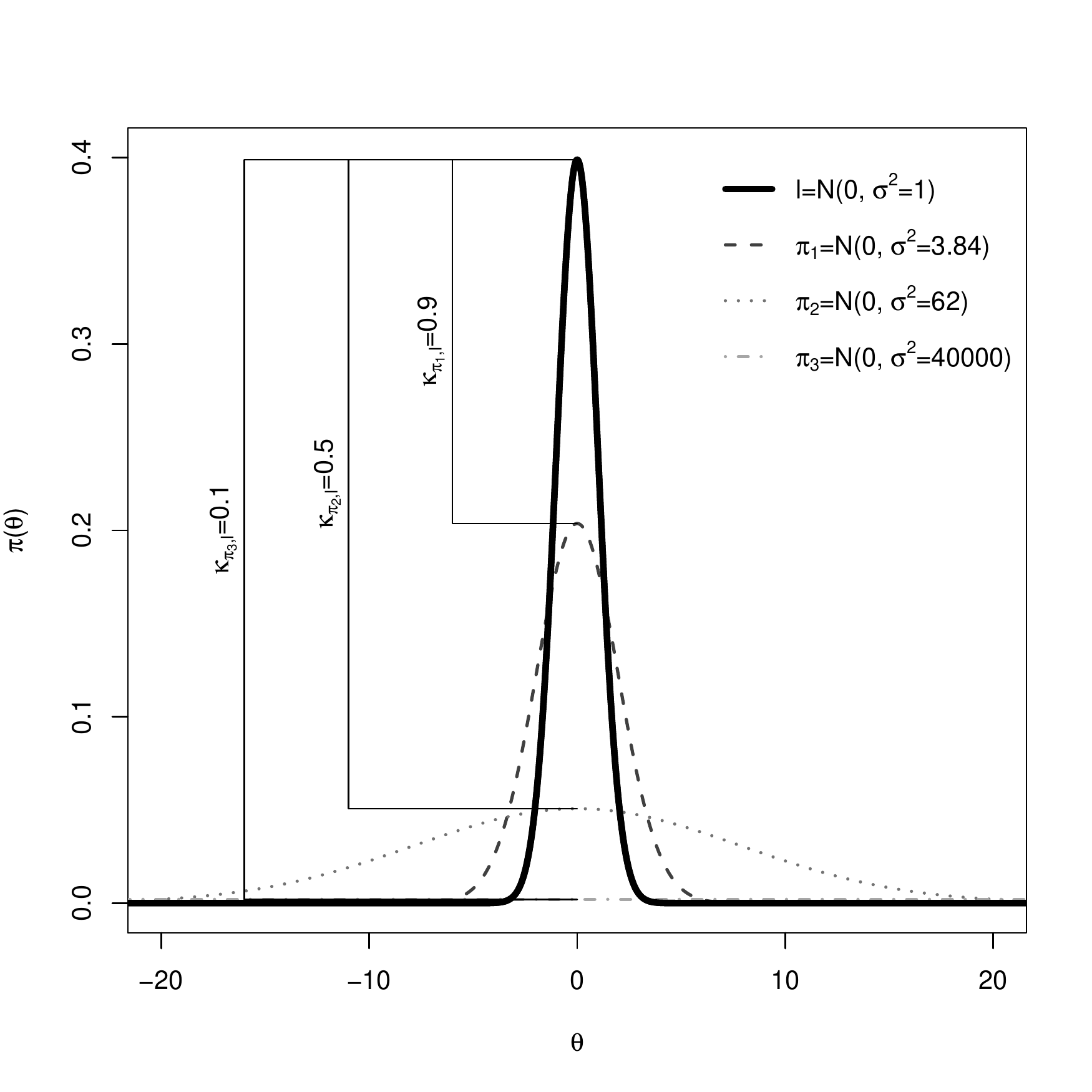} 
  \end{minipage} \\ 
  \begin{minipage}{0.45\linewidth}\centering
    \hspace{-.4cm}\text{\footnotesize{(i)}}    
  \end{minipage}
  \begin{minipage}{0.45\linewidth}\centering
    \hspace{1.25cm}\text{\footnotesize{(ii)}}
  \end{minipage}
\caption{\footnotesize Values of $\kappa_{\pi, \ell}$ when both $\pi$ and $\ell$ are both Gaussian distributions. (i) Gaussian distributions whose means become more separated. (ii) Gaussian distributions that become progressively more diffuse.}
\label{kappa}
\end{figure}

{To build intuition regarding $\kappa_{\pi, \ell}$, we provide Figure~\ref{kappa}, where $\ell$ is set to $\text{N}(0,1)$ while $\pi = \text{N}(m,{\sigma^2})$ varies according to $m$ and ${\sigma^2}$. Figure~\ref{kappa} (i) corresponds to fixing ${\sigma^2}=1$  and varying $m$ while in the right plot $m=0$ is fixed and {$\sigma^2$} varies.  Notice that in plot (i) $\kappa_{\pi, \ell} =0.1$ corresponds to distributions whose means are approximately 3 standard deviations apart while a $\kappa_{\pi, \ell} =0.9$ corresponds to distributions whose means are approximately 0.65 standard deviations apart.  Connecting specific values of $\kappa$ to specific standard deviation distances between means seems like a natural way to quickly get a rough idea of relative differences between two distributions.  In Figure~\ref{kappa} (ii) it appears that if both distributions are centered at the same value, then one distribution must be very disperse relative to the other to produce $\kappa$ values that are small (e.g., $\le 0.1$).  This makes sense as there always exists some mass intersection between the two distributions considered.  
Thus, $\kappa_{\pi,\ell}$---to which we refer as \textit{compatibility}---can be regarded as a measure of the level of agreement between prior and data. Some further comments regarding our geometry are in order:

\begin{itemize}
\item Two different densities $\pi_1$ and $\pi_2$ cannot be collinear: If  $\pi_1=k \pi_2$, then $k=1$, otherwise $\int \pi_2(\btheta) \, \dif \btheta \neq 1$. 
\item A density can be collinear to a likelihood: If the prior is {Uniform then} $p(\btheta \mid {y}) \propto \ell(\btheta)$, and hence the posterior is collinear to the likelihood, i.e., in such a case the posterior simply consists of a renormalization of the likelihood.
\item Two likelihoods can be collinear: Let $\ell$ and $\ell^*$ be the likelihoods based on observing ${y}$ and ${y}^*$, respectively. The strong likelihood principle states that if $\ell(\btheta) = f(\btheta \mid {y}) \propto f(\btheta \mid {y}^*) = \ell^*(\btheta)$, then the \emph{same} inference should be drawn from both samples \citep{BW88}. According to our geometry, this would mean that likelihoods with the same direction yield the same inference.
\end{itemize}
{{As} a  final comment on reparametrizations of the model, interpretations of compatibility should keep a fixed parametrization in mind. That is, we do not recommend comparing prior--likelihood compatibility for models with different parametrizations. Further comments on reparametrizations will be given below in Sections 2.3, 2.4, and 3.2.}

\subsection{Norms and their interpretation}\label{norms}
As $\kappa_{\pi,\ell} $ is comprised of function norms, we dedicate some exposition to how one might interpret these quantities. We start by noting that in some cases the norm of a density is linked to the {variance}, as can be seen in the following example.

\begin{example}\label{unif.normal}\normalfont
Let $U \sim \text{Unif}(a,b)$ and let $\pi({u}) = (b-a)^{-1} I_{(a,b)}({u})$ denote its corresponding density. Then, it holds that $\|\pi\| =
{1 / (12 \sigma_U^2)^{1/4}}$, where the {variance} of $U$ is ${\sigma^2_U = 1 / 12 (b-a)^2}$. Next, consider a Normal model $X \sim \text{N}(\mu,{\sigma^2_X})$ with known {variance $\sigma^2_X$} and let $\phi$ denote its corresponding density. It can be shown that $\|\phi\|=\{\int_{\mathbb{R}} \phi^2(x; \mu, {\sigma^2_X}) \,\dif \mu\}^{1/2} = {1/(4 \pi \sigma_X^2)^{1/4}}$ which is a function of ${\sigma_X^2}$. \end{example}

The following proposition explores {how the norm of a general prior density, $\pi$, relates with that of a Uniform density, $\pi_0$.}

\begin{proposition} \label{NormInterpretation}
  Let $\Theta \subset \mathbb{R}^{p}$ with $\lambda(\Theta)<\infty$ where $\lambda$ denotes the Lebesgue measure. Consider $\pi$ : $\Theta\to [0, \infty)$ a probability density with $\pi \in L_2(\Theta)$ and let $\pi_{0}$ denote a Uniform density on $\Theta$, then
\begin{align}\label{pi_0}
  \|\pi\|^{2}   =\|\pi-\pi_{0}\|^{2}+\|\pi_{0}\|^{2}.
\end{align}
\end{proposition}
\noindent Since $\| \pi_{0} \|^{2}$  is constant, $\|\pi\|^{2}$ increases as $\pi$'s mass becomes more concentrated (or less Uniform). Thus, as can be seen from \eqref{pi_0}, $\|\pi\|$ is a measure of how much $\pi$ differs from a Uniform distribution over $\Theta$. This interpretation cannot be applied to $\Theta$'s that do not have finite Lebesgue measure as there is no corresponding proper Uniform distribution. Nonetheless, the notion that the norm of a density is a measure of its peakedness may be applied whether or not $\Theta$ has finite Lebesgue measure. {To see this, evaluate $\pi(\theta)$ on a grid $\theta_1 < \cdots < \theta_D$ and consider the vector $p = (\pi_1, \dots, \pi_D)$, with $\pi_d = \pi(\theta_d)$ for $d = 1, \ldots, D$. 
The larger the norm of the vector $p$, the higher the indication that certain  components would be far from the origin---that is, $\pi(\theta)$ would be peaking for certain $\theta$ in the grid. Now, think of a density as a vector with infinitely many components (its value at each point of the support) and replace summation by integration to get the $L_2$ norm.}
Therefore, $\| \cdot \|$ can be used to compare the `informativeness' of two competing priors with $\| \pi_1 \| < \| \pi_2 \|$ indicating that $\pi_1$ is less informative. 

Further reinforcing the idea that the norm is related to the peakedness of a distribution, there is an interesting connection between $\| \pi \|$ and the (differential) entropy (denoted by $H_{\pi}$)  which is described in the following {proposition}. 

\begin{proposition}\label{expansion}
Suppose $\pi \in L_2(\Theta)$ is a continuous density on a compact $\Theta \subset \mathbb{R}^p$, and that  $\pi(\btheta)$ is differentiable on $\emph{int}(\Theta)$. Let $H_{\pi}=-\int_{\Theta} \pi({\theta}) \log \pi({\theta}) \, \emph{d}{\theta}$. Then, it holds that  
\begin{equation}
  \label{connection}
  \|\pi\|^2 = 1-H_{\pi}+o\{\pi(\btheta^*)-1\},
\end{equation}
for some $\btheta^* \in \emph{int}(\Theta)$.
\end{proposition}

The expansion in \eqref{connection} hints that the norm of a density and the entropy should  be negatively related, and hence as the norm of a density increases,  its mass becomes more concentrated. In terms of priors, this suggests that priors with a large norm should be more `peaked' relative to priors with a smaller norm. Therefore, the magnitude of a prior appears to be linked to its peakedness (as is demonstrated in \eqref{pi_0} and in Example~\ref{unif.normal}). 
While this might also be viewed as `informativeness,' the $\text{Beta}(a,b)$ density has a higher norm if $(a,b)\in(1/2,1)^2$ than if $a=b=1$, possibly placing this interpretation at odds with the notion that  $a$ and $b$ represent `prior successes' and `prior failures' in the Beta--Binomial setting.   {As will be further discussed in Section~\ref{expf}, a reviewer recognized that this seeming paradox is a consequence of the parameterization employed and is avoided when using the log-odds as the parameter.}

As can be seen from \eqref{connection}, the connection between entropy and $\| \pi \|$ is an approximation at best. Just as a first-order Taylor expansion provides a poor polynomial approximation for points that are far from the point under which the expansion is made, the expansion in \eqref{connection} will provide a poor entropy approximation when  $\pi$ is not similar to a standard Uniform-like distribution $\pi_0$. However, since $\|\pi_0\|^2 = 1-H_{\pi_0}$, the approximation is exact for a standard Uniform-like distribution. We end this discussion by noting that integrals related to $\| \pi \|^2$ also appear in physical models on $L_2$-spaces and they are usually interpreted as the total energy of a physical system \citep[][p.~142]{HN05}, and there is {considerable} frequentist literature on the estimation of {the integrated square of a density}  \citep[see][and references therein]{GN08}. Now, to illustrate the information that $\| \cdot \|$ and $\kappa$ provide, we consider the example described in Section~\ref{introduction}.

\begin{figure}\centering
  \begin{minipage}[c]{0.45\textwidth}  \centering
    \includegraphics[height = 6.3cm]{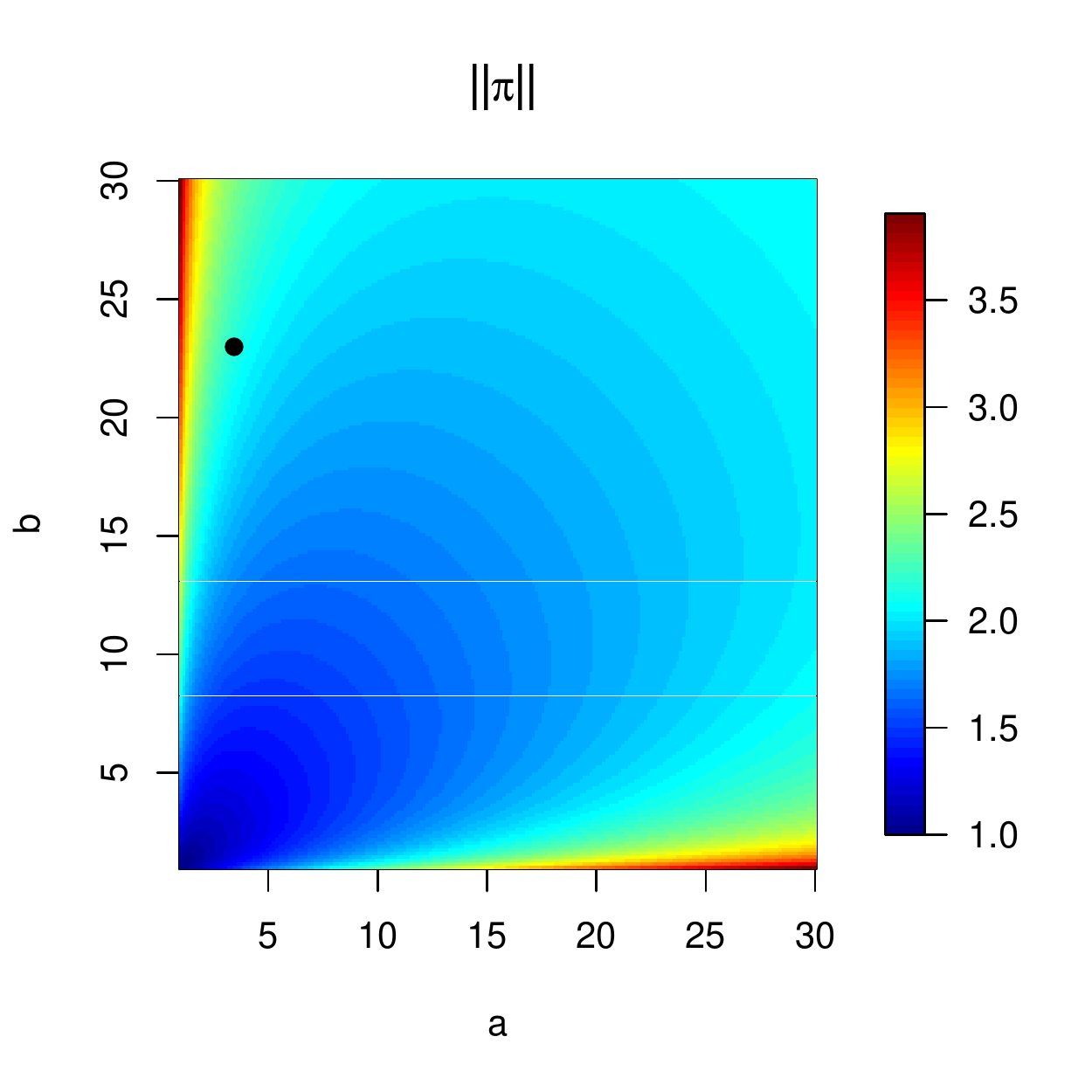} 
  \end{minipage} 
  \begin{minipage}[c]{0.45\textwidth}   \centering
    \includegraphics[height = 6.3cm]{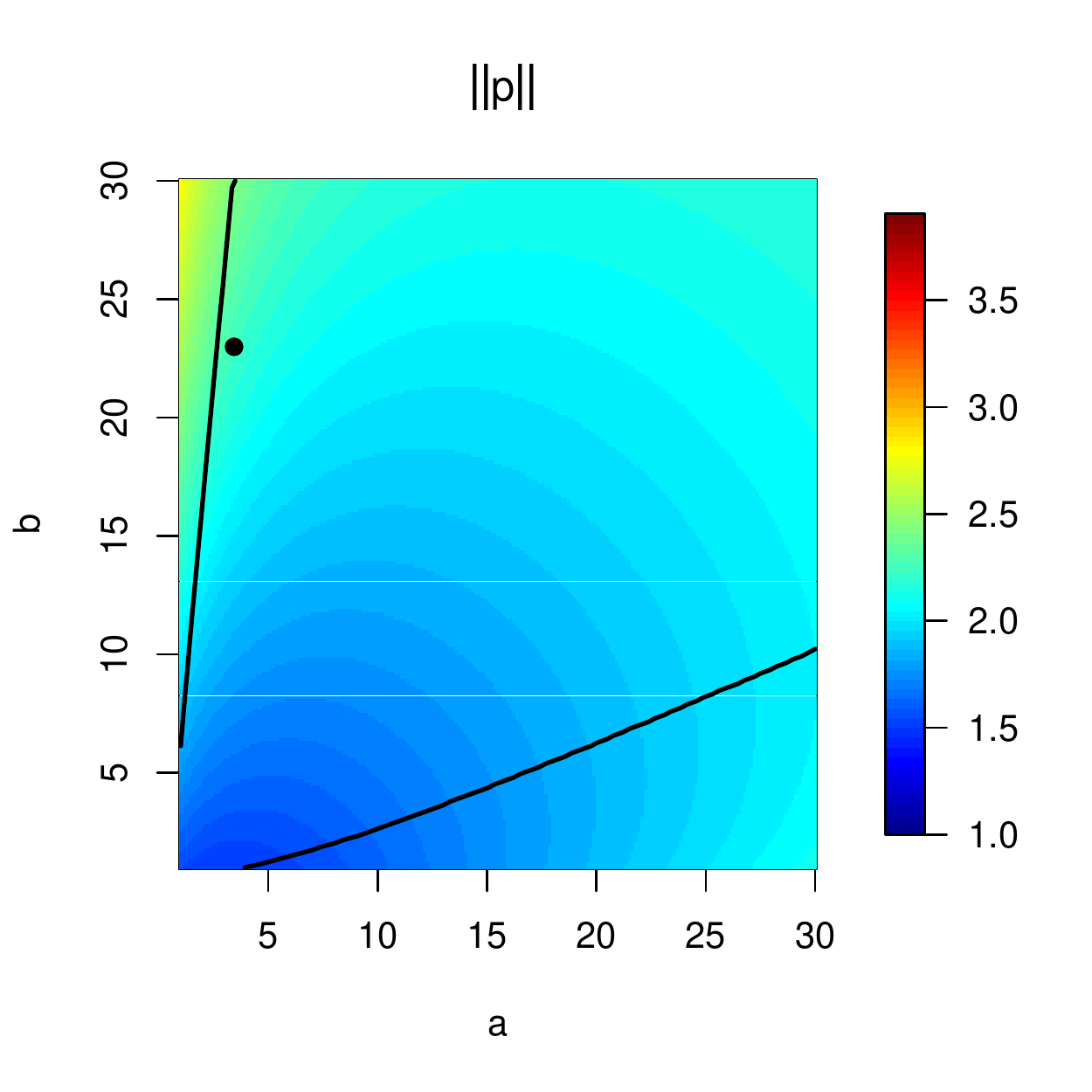} 
  \end{minipage} \\ 
  \begin{minipage}[c]{0.45\textwidth} \hspace{2.2cm}
    \hspace{1cm}\text{\footnotesize{(i)}}
  \end{minipage}
  \begin{minipage}[c]{0.45\textwidth} \hspace{2.3cm}
    \hspace{1cm}\text{\footnotesize{(ii)}}
\end{minipage}
\caption{\footnotesize Prior and posterior norms for on-the-job drug usage toy example. Contour plots depicting the $\|\cdot\|$ associated with a Beta$(a,b)$ prior (i) and the corresponding {Beta$(a^{\star},b^{\star})$} posterior (ii), with $a^{\star} =  a + 2$ and $b^{\star} = b + 8$. Solid lines in (ii) indicate boundaries delimiting the region of values of $a$ and $b$ for which $\|\pi\| > \|p\|$. The solid dot ($\bullet$) corresponds to $(a,b)= (3.44, 22.99)$ (values employed by {\citealt[][pp.~26--27]{CAL11}}).}
\label{norms}
\end{figure}

\begin{example}[On-the-job drug usage toy example, cont.~1]\label{ex2}\normalfont
From the example in the Introduction we have $\theta \mid {{y}} \sim \text{Beta} (a^{\star}, b^{\star})$ with $a^{\star} = n_1 + a = 2 + a$ and $b^{\star} = n - n_1 + b = 8 + b$. The norm of the prior, posterior, and likelihood are respectively given by
\begin{equation}\label{normm}
    \| \pi(a,b) \| = \frac{\{B(2a-1,2b-1)\}^{1/2}}{B(a,b)},  
\end{equation}
and $\| p(a,b) \| =  \| \pi(a^{\star}, b^{\star}) \|$, 
with $a,b > 1/2$, and 
\begin{equation*}
  \|\ell\| = {\binom{n}{n_1}} \{B\left(2 n_1 + 1, 2 \left(n- n_1\right)+1\right)\}^{1/2},
\end{equation*}
where $B(a,b) = \int_0^1 u^{a-1} (1-u)^{b-1} \, \dif u$.

Figure~\ref{norms} (i) plots $ \| \pi(a,b) \|$ and Figure~\ref{norms} (ii) plots $ \| p(a,b) \|$ as functions of $a$ and $b$. We highlight the prior values  $(a_0,b_0)= (3.44, 22.99)$ which were employed by \cite{CAL11}.  Because prior densities with large norms will be more peaked relative to priors with small norms,  $\|\pi(a_0, b_0)\|=2.17$ is more peaked than $\|\pi(1,1)\|=1$ (Uniform prior) {indicating that $\|\pi(a_0, b_0)\|$ is more `informative' than $\|\pi(1,1)\|$}.  The norm of the posterior for these same pairs is $\|p(a_0, b_0)\|=2.24$ and $\|p(1,1)\|=1.55$, meaning that the posteriors will have mass more concentrated than the corresponding priors.  The lines found in Figure~\ref{norms} (ii) represent boundary lines such that all $(a,b)$ pairs that fall outside of the boundary produce $\| \pi(a,b) \| > \| p(a,b) \|$ which indicates that the prior is more peaked than the posterior (typically an undesirable result).    If we used an extremely peaked prior, say $(a_1,b_1)= (40, 300)$, then we would get $\|\pi(a_1,b_1)\|=4.03$ and $\|p(40,300)\|=4.04$ indicating that the peakedness of the prior and posterior densities is essentially the same. 

Considering $\kappa_{\pi,\ell}$, it follows that 
\begin{equation}
    \kappa_{\pi,\ell}(a,b) =  \frac{B(a^{\star}, b^{\star})}{\{B(2a-1, 2b-1) B(2 n_1 + 1, 2(n-n_1 ) + 1)\}^{1/2}},
 \label{priorlikcomp}
\end{equation}
{with $a^{\star} = n_1 + a$ and $b^{\star} = n - n_1 + b$.} Figure~\ref{KappaPrPoKappPrPr} (i) plots values of $\kappa$ as a function of prior parameters $a$ and $b$ with $\kappa_{\pi, \ell}(a_0, b_0) \approx 0.69$ being highlighted indicating a great deal of agreement with the likelihood.   In this example a lack of prior--data compatibility would occur (e.g., $\kappa_{\pi, \ell} \le 0.1$) for priors that are very peaked at $\theta > 0.95$ or for priors that place substantial mass at {$\theta < 0.05$.}

The values of the hyperparameters $(a,b)$ which, according to $\kappa_{\pi,\ell}$, are more compatible with the data (i.e., those that maximise $\kappa$) are given by $(a^*, b^*)=(3,9)$ and are highlighted with a star (\textbf{\textasteriskcentered}) in Figure~\ref{KappaPrPoKappPrPr} (i). In Section~\ref{maxcompatible} we provide some connections between this prior and maximum likelihood estimators.
\end{example}

\begin{figure}[h]
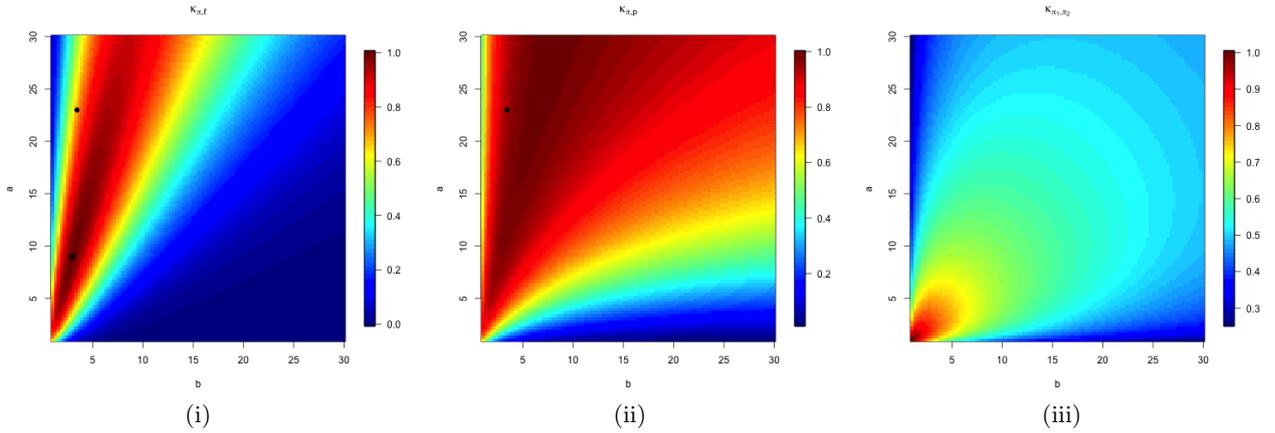

  \centering
  \begin{minipage}{0.3\linewidth} 
    \includegraphics[height = 5.5cm]{BetaBernKappa3.pdf} 
  \end{minipage} \hspace{.3cm}
  \begin{minipage}[c]{0.3\linewidth}
    \includegraphics[height = 5.5cm]{BetaBernKappip2.pdf} 
  \end{minipage} \hspace{.3cm}
  \begin{minipage}[c]{0.3\linewidth}
    \includegraphics[height = 5.5cm]{BetaBernKappi1pi3.pdf} 
  \end{minipage} \\ 
  \hspace{-.3cm}
  \begin{minipage}{0.33\linewidth} 
    \hspace{2.7cm}
    \text{\footnotesize{(i)}}
  \end{minipage}
  \begin{minipage}{0.33\linewidth} 
    \hspace{2.6cm}
    \text{\footnotesize{(ii)}}
  \end{minipage}
  \begin{minipage}{0.33\linewidth} 
    \hspace{2.5cm}
    \text{\footnotesize{(iii)}}
  \end{minipage}
\caption{\footnotesize Compatibility ($\kappa$) for on-the-job drug usage toy illustration as found in \eqref{priorlikcomp} and Example~4. (i) Prior--likelihood compatibility,  $\kappa_{\pi, \ell}(a, b)$; the black star (\textbf{\textasteriskcentered}) corresponds to $(a^*, b^*)$ which maximise $\kappa_{\pi, \ell}(a, b)$. (ii) Prior--posterior compatibility, $\kappa_{\pi, p}(a,b)$. (iii) Prior--prior compatibility, $\kappa_{\pi_1, \pi_2}(1, 1, a, b)$, where $\pi_1 \sim \text{Beta}(1,1)$ and $\pi_2 \sim \text{Beta}(a,b)$. In (i) and (ii) the solid dot ($\bullet$) corresponds to $(a,b)= (3.44, 22.99)$ (values employed by {\citealt[][pp.~26--27]{CAL11}}).}
\label{KappaPrPoKappPrPr}
\end{figure}

\subsection{Angles between other vectors}\label{angles.norms}
As mentioned, we are not restricted to use $\kappa$ only to compare $\pi$ and $\ell$. Angles between densities, and between likelihoods and densities or even between two likelihoods are available. We explore these options further using the example provided in the Introduction.

\begin{example}[On-the-job drug usage toy example, cont.~2]\normalfont \label{exkpip}
Extending Example~\ref{ex2} and \eqref{priorlikcomp} we calculate
\begin{equation*}
  \kappa_{\pi,p}(a, b) = 
  \frac{B({a + a^{\star} - 1, b + b^{\star} - 1})}{\{B(2a-1,2b-1) 
    B({2a^{\star}}-1, {2b^{\star}}-1)\}^{1/2}}, 
\end{equation*}
{with $a^{\star} = n_1 + a$ and $b^{\star} = n - n_1 + b$;} for $\pi_1 \sim \text{Beta}(a_1, b_1)$ and $\pi_2 \sim  \text{Beta}(a_2, b_2)$,
\begin{equation*}
\begin{split}
  \kappa_{\pi_1, \pi_2}(a_1, b_1, a_2, b_2) = 
  \frac{B(a_1 + a_2 -1, b_1 + b_2 - 1)}{\{B(2a_1 - 1,2b_1 - 1) B(2a_2 - 1, 2b_2 - 1)\}^{1/2}}.
\end{split}
\end{equation*}
To visualize how the hyperparameters influence $\kappa_{\pi, p}$ and $\kappa_{\pi_1, \pi_2}$ we provide Figures~\ref{KappaPrPoKappPrPr} (ii) and (iii). Figure~\ref{KappaPrPoKappPrPr} (ii) again highlights the prior used in \cite{{CAL11}} with $\kappa_{\pi, p}(a_0, b_0) \approx 0.95$; see solid dot ($\bullet$). This value of $\kappa_{\pi, p}$ implies that both prior and posterior are concentrated on essentially the same subset of $[0,1]$,  indicating a large amount of agreement between them. Disagreement between prior and posterior takes place with priors concentrated on high probabilities of $\theta$ being greater than 0.8. In Figure~\ref{KappaPrPoKappPrPr} (iii), $\kappa_{\pi_1, \pi_2}$ is largest when $\pi_2$ is close to $\text{Unif}(0,1)$ (the distribution of $\pi_1$) and gradually drops off as $\pi_2$ becomes more peaked and/or less symmetric. 
\end{example}
In the next example, we use another data illustration to demonstrate the application of $\kappa$ to a two-parameter model.
\begin{example}[Midge wing length data]\normalfont
Let $Y_1, \dots, Y_n \mid \mu, \sigma^2 \overset{\text{iid}}{\sim} \text{N}(\mu, \sigma^2)$, and $\mu \mid \sigma^2 \sim \text{N}(\mu_0, \sigma^2/\eta_0)$ and $\sigma^2 \sim \text{IG}(\nu_0/2,\sigma_0^2\nu_0/2)$; we refer to this conjugate prior distribution as $\text{NIG}(\mu_0,\eta_0,\nu_0,\sigma_0^2)$. In comparing $\pi_1 = \text{NIG}(\mu_1,\eta_1,\nu_1,\sigma_1^2)$ and $\pi_2 = \text{NIG}(\mu_2,\eta_2,\nu_2,\sigma_2^2)$, $\kappa_{\pi_1, \pi_2}$ may be expressed as,
\begin{equation}
  \kappa_{\pi_1,\pi_2} = \frac{(\pi_A \pi_B)^{1/2}}{\pi_C}\Bigr|_{\mu=0,\sigma^2=1},
 \label{kappa_pi1pi2}
\end{equation}
with 
\begin{equation*}
\begin{split}
  \pi_A &= \text{NIG}(\mu_1,2\eta_1,2\nu_1+3,\nu_1\sigma_1^2/(\nu_1+3/2)), \quad 
  \pi_B = \text{NIG}(\mu_2,2\eta_2,2\nu_2+3,\nu_2\sigma_2^2/(\nu_2+3/2)), \quad \\
  \pi_C &= \text{NIG}((\eta_1\mu_1+\eta_2\mu_2)/(\eta_1+\eta_2), \eta_1+\eta_2,\nu_1+\nu_2+3, \\ 
  & \hspace{1.3cm}\{\nu_1\sigma_1^2+\nu_2\sigma_2^2 +\eta_1\eta_2(\mu_1-\mu_2)^2/(\eta_1+\eta_2)\}/(\nu_1+\nu_2+3)).
\end{split}
\end{equation*}
\begin{figure}  \centering
  \begin{minipage}[c]{0.45\textwidth}  \centering
    \includegraphics[height = 6.3cm]{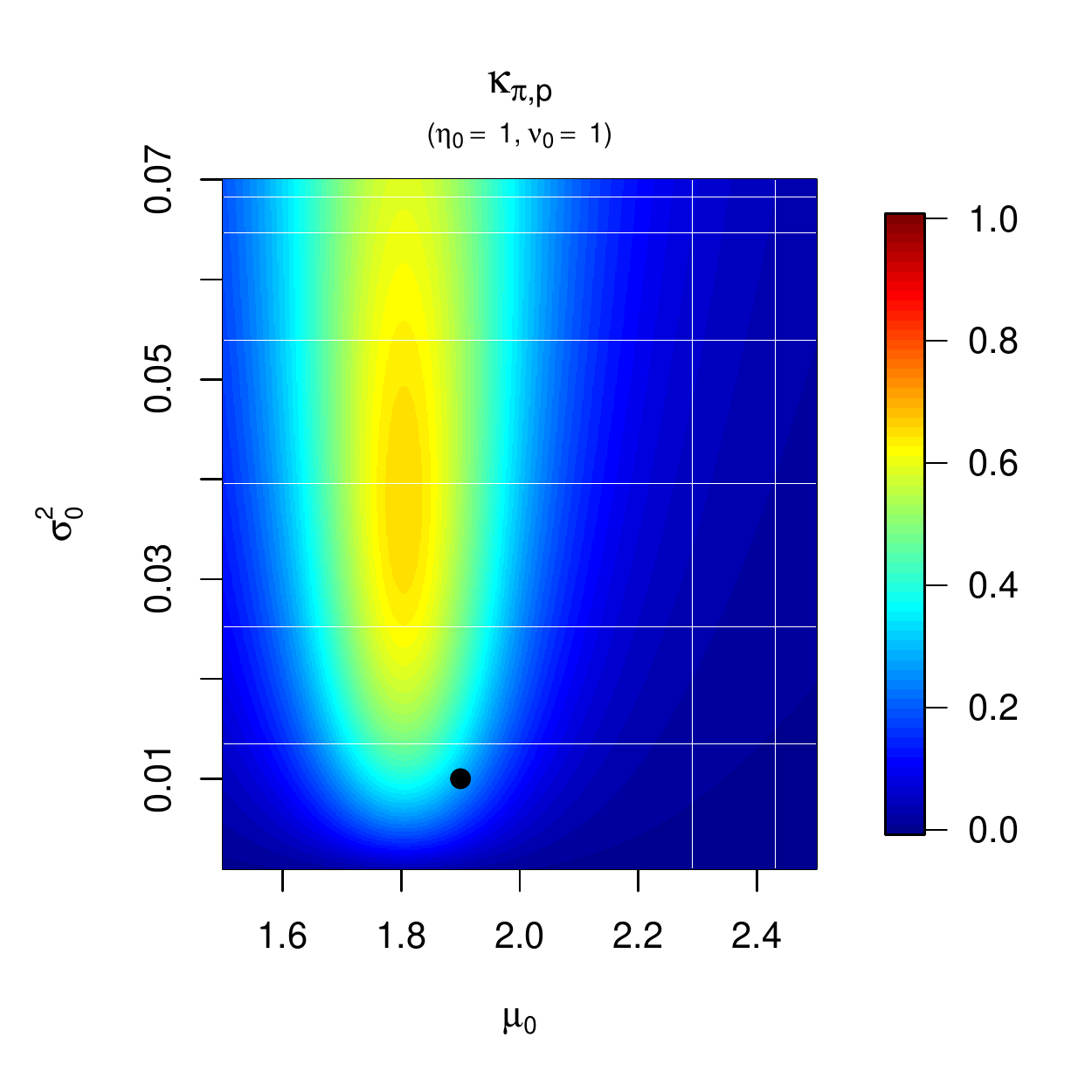} 
  \end{minipage} 
  \begin{minipage}[c]{0.45\textwidth}  \centering
    \includegraphics[height = 6.3cm]{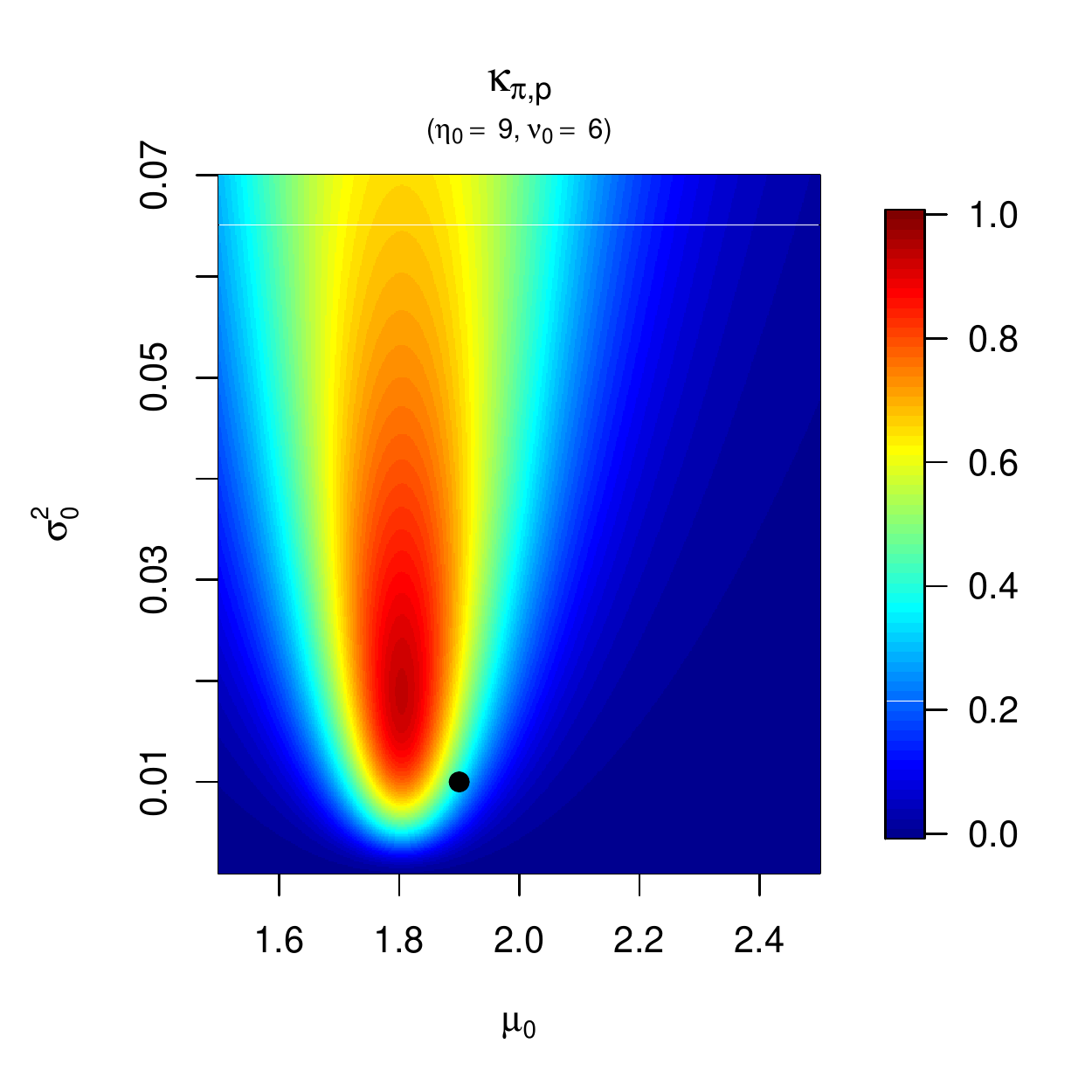} 
  \end{minipage} \vspace{-.15cm}\\
  \begin{minipage}[c]{0.45\textwidth} 
    \hspace{3.3cm}
    \text{\footnotesize{(i)}}
  \end{minipage} 
  \begin{minipage}[c]{0.45\textwidth}  
    \hspace{3.3cm}
    \text{\footnotesize{(ii)}}
  \end{minipage}
\caption{\footnotesize Prior--posterior compatibility, $\kappa_{\pi, p}(\mu_0, \eta_0,\nu_0,\sigma_0^2)$, for midge wing lengths data from Example~\ref{NIG}. In (i) $\eta_0$ and $\nu_0$ are fixed at one, whereas in (ii) $\eta_0$ is fixed at nine and $\nu_0$ is fixed at six. The solid dot ($\bullet$) corresponds to $(\mu_0, \sigma_0^2) = (1.9, 0.01)$ which is here used as a baseline given that hyperparameters employed by  \citet[][pp.~72--76]{Hoff2009} are $\mu_0=1.9, \eta_0=1, \nu_0=1$, and $\sigma_0^2=0.01$.}
\label{NNIGpriPOS}
\end{figure}
Note that \eqref{kappa_pi1pi2} {(whose derivation can be found in Section~5.1 of the Supplementary Materials)} may also be used to compute $\kappa_{\pi_1, p}$, since $p = \text{NIG}(\mu^\star, \eta^\star, \nu^\star, \sigma^{2\star})$, with
\begin{equation*}
  \begin{cases}
    \begin{split}
      \mu^\star &= (n\bar{Y}+\eta_0\mu_0)/(n+\eta_0), \quad 
      \eta^\star = \eta_0+n, \quad 
      \nu^\star = \nu_0 + n, \\
      \sigma^{2\star} &= \left\{\nu_0\sigma_0^2 + \sum_{i = 1}^n{(Y_i-\bar{Y})^2} +
      \eta_0 n {(\eta^\star)}^{-1}(\mu_0-\bar{Y})^2\right\}/\nu^\star.
    \end{split}
  \end{cases}
\end{equation*}
Computation of $\kappa_{\pi_1, \ell}$ also adheres to Equation~\eqref{kappa_pi1pi2} if $n>3$ and $\pi_2 =\text{NIG}(\bar{Y}, n, n-3,  \sum_{i = 1}^n{(Y_i-\bar{Y})^2}/(n-3))$ because then $\ell$ is collinear to $\pi_2$.
\citet[][pp.~72--76]{Hoff2009} applied this model to a dataset of nine midge wing lengths, where he set $\mu_0=1.9$, $\eta_0=1$, $\nu_0=1$, and $\sigma_0^2=0.01$, while $\bar{Y}=1.804$ and $\sum_{i = 1}^n{(Y_i-\bar{Y})^2}\approx 0.135$. This yields  $\kappa_{\pi,p}\approx 0.28$, and thus the agreement between the prior and posterior is not particularly strong. Figure~\ref{NNIGpriPOS} (i) displays $\kappa_{\pi, p}$, as a function of $\mu_0$ and $\sigma_0^2$ while fixing $\nu_0=1$ and $\eta_0=1$. To evaluate how $\kappa_{\pi, p}$ is affected by $\nu_0$ and $\eta_0$, the analogous plot is displayed as Figure~\ref{NNIGpriPOS} (ii) when these values are fixed at $\nu_0=6$ and $\eta_0=9$; these alternative values for $\nu_0$ and $\eta_0$ are those which allow the compatibility between the prior and likelihood to be maximised. It is apparent from Figure~\ref{NNIGpriPOS} that a larger $\sigma_0^2$ increases $\kappa_{\pi,p}$ substantially, and a simultaneous increase of $\nu_0$ and $\eta_0$ would further propel this increase.
\label{NIG}
\end{example}
{Some comments on reparametrizations are in order. We focus on the case of compatibility between two priors with a single parameter, but the rationale below also applies to compatibility between a prior and posterior, and in multiparameter settings. Let $\theta_1 \sim \pi_1$ and $\theta_2 \sim \pi_2$; further, let $g(\theta) = \lambda$ be a monotone increasing function, with range $\Lambda$, and let 
    \begin{equation*}
      \pi^g_1(\lambda) = \frac{\pi_1(g^{-1}(\lambda))}{g'(g^{-1}(\lambda))},
      \quad \pi^g_2(\lambda) = \frac{\pi_2(g^{-1}(\lambda))}{g'(g^{-1}(\lambda))},      
    \end{equation*}
    be prior densities of the transformed parameters, $g(\theta_1)$ and $g(\theta_2)$. It thus follows that 
    \begin{equation*}
      \begin{split}
       \frac{\int_{\Lambda} \pi^g_1(\lambda) \pi^g_2(\lambda) \,\dif \lambda}{[\int_{\Lambda} \{\pi^g_1(\lambda)\}^2 \dif \lambda  \int_{\Lambda} \{\pi^g_2(\lambda)\}^2 \dif \lambda]^{1/2}} 
=\frac{\int_{\Theta} \pi_1(\theta) \pi_2(\theta) / g'(\theta) \, \dif \theta}{[\int_{\Theta} \{\pi_1(\theta)\}^2 / g'(\theta) \, \dif \theta \,  \int_{\Theta} \{\pi_2(\theta)\}^2 / g'(\theta) \, \dif \theta]^{1/2}}.
      \end{split}
    \end{equation*}
The version of compatibility discussed in this section is thus invariant to linear transformations of the parameter. A variant to be discussed in Section~\ref{ac} is more generally invariant to monotone increasing transformations.}

\subsection{Max-compatible priors and maximum likelihood estimators}\label{maxcompatible}
In Example~\ref{ex2}, we briefly alluded to a connection between priors maximising prior--likelihood compatibility $\kappa_{\pi, \ell}$ (to be termed as max-compatible priors) and maximum likelihood (ML) estimators, on which we now elaborate. Below, we use the notation $\pi(\btheta \mid \balpha)$ to denote a prior on $\btheta \in \Theta$, with $\balpha \in {A}$ are hyperparameters, and where $\dim({A}) = q$ and $\dim(\Theta) = p$. (Think of the Beta--Binomial model, where $\theta \in \Theta = (0,1)$, and $\balpha = (a, b) \in {A} = (0, \infty)^2$.) 

\begin{definition}[Max-compatible prior]
Let ${y} \sim f(\, \cdot \mid \btheta)$, and let $\mathcal{P} = \{\pi(\btheta \mid \balpha): \balpha \in {A}\}$ be a family of priors for $\btheta$. If there exists $\balpha^*_{{y}} \in {A}$, such that $\kappa_{\pi, \ell} (\balpha^*_{{y}}) = 1$, the prior $\pi(\btheta \mid  \balpha^*_{{y}}) \in \mathcal{P}$ is said to be \textit{max-compatible}, and $\balpha^*_{{y}}$ is said to be a max-compatible hyperparameter.
\label{max-compatibility}
\end{definition}
\noindent The max-compatible hyperparameter, $\balpha^*_{{y}}$, is by definition a random vector, and thus a max-compatible prior density is a random function. Geometrically, a prior is max-compatible if and only if it is collinear to the likelihood in the sense that $\kappa_{\pi, \ell} (\balpha^*_{{y}}) = 1$ if and only if $\pi(\btheta \mid \balpha_{{y}}^*) \propto f({y} \mid \btheta)$, for all $\btheta \in \Theta$. 

The following example suggests there could be a connection between the ML estimator of $\btheta$ and the max-compatibility parameter $\balpha^*_{{y}}$.

\begin{example}[Beta--Binomial]\normalfont \label{bbin}
  Let $n_1 \mid \theta \sim \text{Bin}(n, \theta)$, and suppose $\theta \sim \text{Beta}(a, b)$. Here, $\mathcal{P} = \{\beta(\theta \mid a, b): (a,b) \in (1/2,\infty)^2\},$
  with 
  $\beta(\theta \mid a, b) = \theta^{a - 1} (1 - \theta)^{b - 1} / B(a, b).$
  It can be shown that the max-compatible prior is $\pi(\theta \mid a^*, b^*) = \beta(\theta \mid a^*, b^*)$, where $a^* = 1 + n_1$, and $b^* = 1 + n - n_1$, so that
  \begin{equation}\label{m1}
    \begin{split}  
    \widehat{\theta} = \arg \max_{\theta \in (0,1)} f({n_1} \mid \theta) 
    = {\frac{n_1}{n}} = \frac{a^{*} - 1}{a^{*} + b^{*} - 2} =: m(a^{*}, b^{*}),
    \end{split}
  \end{equation}
{with $f(n_1 \mid \theta) = \binom{n_1}{n} \theta^{n_1} (1 - \theta)^{n - n_1}$.}
\label{beta-binomial.ex}
\end{example}

\noindent A natural question is whether there always exists a function {$m: {A} \to \Theta$}, as in \eqref{m1}, linking the max-compatible parameter with the ML estimator? The following theorem addresses this. 
\begin{proposition}\label{max-thm}
  Let ${y} \sim f(\,\cdot \mid \btheta)$, and  {let $\widehat{\btheta}$ be the ML estimator of $\btheta$}. In addition, let $\mathcal{P} = \{\pi(\btheta \mid \balpha): \balpha \in {A}\}$ be a family of priors for $\btheta$. If there exists a unimodal max-compatible prior, then 
  \begin{equation*}
    \begin{split}
    \widehat{\btheta} = \arg \max_{\btheta \in \Theta} f({y} \mid \btheta) 
    = m_{\pi}(\balpha^*_{{y}}) := \arg \max_{\btheta \in \Theta} \, \pi (\btheta \mid \balpha^*_{{y}}).
  \end{split}
  \label{max.connect}
  \end{equation*} 
\end{proposition}
\noindent Proposition~\ref{max-thm} states that the mode of the max-compatible prior coincides with the ML estimator, and in Example~\ref{bbin},  $m(a^*,b^*) = (a^{*} - 1) / (a^{*} + b^{*} - 2)$ is indeed the mode of a Beta prior. {A comment on parametrizations is in order. A corollary to Proposition~\ref{max-thm} is that, due to invariance of ML estimators, if $m_{\pi}(\balpha^*_{{y}})$ is the mode of the max-compatible prior for $\theta$ and $g(\theta) = \lambda$ is a function, then $g(m_{\pi}(\balpha^*_{{y}}))$ is the mode of the max-compatible prior of the transformed parameter $\pi^{g}(\lambda \mid \alpha^*_y)$. Formally, 
  \begin{equation*}
    g(\widehat \theta) = \widehat{\lambda} = \arg \max_{\lambda \in \Lambda} \sup_{\theta \in \Theta_\lambda} f({y} \mid \btheta) 
    = g(m_{\pi}(\balpha^*_{{y}})) = \arg \max_{\lambda \in \Lambda} \, \pi^g(\lambda \mid \balpha^*_{{y}}), 
  \end{equation*}
with $\Theta_\lambda = \{\theta: g(\theta) = \lambda\}$ and where $\Lambda$ is the range of $g$.
} 

{The max-compatible prior is a `prior' to the extent that it belongs to a family of priors, but it is basically a posterior distribution (it depends on the data). Also, there are some links between the max-compatible prior and  {Hartigan}'s maximum likelihood prior \citep{hartigan1998}, which will be  clarified in Section~\ref{expf}.}


\subsection{Compatibility in the exponential family}\label{expf}
{We now consider compatibility in the exponential family with density 
  \begin{equation*}
    f_\theta(y) = h(y) \exp\{\eta_\theta^{\T} T(y) - A(\eta_\theta)\},
  \end{equation*}
for given functions $T$ and $h$, and with $A(\eta_\theta) = \log [\int h(y) \exp\{\eta_{\theta}^{\T} T(y)\}\, \dif y] < \infty$ denoting the so-called cumulant function. Given a random sample from an exponential family, $Y_1, \ldots, Y_n \mid \theta \iid f_\theta$, it follows that 
\begin{equation*}
  \ell(\theta) = \bigg[\prod_{i = 1}^n h(Y_i)\bigg] \exp\bigg\{\eta_{\theta}^{\T} \sum_{i = 1}^n T(Y_i) - n A(\eta_{\theta})\bigg\}.
\end{equation*}
The conjugate prior is known to be 
\begin{equation}\label{pifam}
 \pi(\theta \mid \tau, n_0) = K(\tau, n_0) \exp\{\tau^{\T} \eta_{\theta} - n_0 A(\eta_{\theta})\}, 
\end{equation}
where $\tau$ and $n_0$ are parameters, and
\begin{equation}\label{K}
K(\tau, n_0) = \bigg[ 
  \int_{\Theta} \exp\{\tau^{\T} \eta_\theta - n_0 A(\eta_\theta)\} \, 
  \dif \theta\bigg]^{-1}.
\end{equation}
The posterior density is $\pi(\theta \mid \tau + \sum_{i = 1}^nT(Y_i), n_0 + n)$, with $\pi(\theta \mid \tau, n_0)$ defined as in \eqref{pifam}; cf \cite{diaconis1979}. In this context, compatibility can be expressed using normalizing constants from various members of the conjugate prior family as follows
\begin{equation}
  \begin{cases}\displaystyle
    \kappa_{\pi, \ell}(\tau, n_0) = \frac{\{K(2 \tau, 2n_0) K(2 \sum_{i = 1}^n T(Y_i), 2n)\}^{1/2}}{K(\tau + \sum_{i = 1}^n T(Y_i), n_0 + n)}, \\ \displaystyle
    \kappa_{\pi, p}(\tau, n_0) =  \frac{\{K(2 \tau, 2n_0) K(2 \{\tau + \sum_{i = 1}^n T(Y_i)\}, 2\{n_0 + n\})\}^{1/2}}{K(2\tau + \sum_{i = 1}^n T(Y_i), 2n_0 + n)}, \\ \displaystyle
    \kappa_{p, \ell}(\tau, n_0) = \frac{\{K(2 \{\tau + \sum_{i = 1}^n T(Y_i)\}, 2\{n_0 + n\}) K(2\sum_{i = 1}^n T(Y_i), 2n)\}^{1/2}}{K(\tau + 2 \sum_{i = 1}^n T(Y_i), n_0 + 2n)},
  \end{cases}
\label{kapfam}
\end{equation}
for $(\tau, n_0)$ for which the normalizing constants in \eqref{kapfam} are defined. The max-compatible prior in the exponential family is given by the following data-dependent prior
\begin{equation}\label{maxmax}
  \pi\bigg(\theta \mid \sum_{i = 1}^nT(Y_i), n\bigg),
\end{equation}
 with $\pi(\theta \mid \tau, n)$ as in \eqref{pifam}. Special cases of the results in \eqref{kapfam} and \eqref{maxmax} were manifest for instance in \eqref{priorlikcomp}, Example~\ref{exkpip}, and Example~\ref{bbin}. 
}

As pointed out by a reviewer, working with the canonical parametrization brings  numerous advantages, especially when measuring compatibility. Since the parametrization of a model is arbitrary (and hence the interpretation of the parameter may be different for each model) it is desirable to work in terms of a parametrization that preserves the same meaning regardless of the model under consideration. For exponential families, a natural choice is the canonical parameter $\eta_{\theta} = \theta$. For one thing, the conjugate prior on the canonical parameter always exists under very general conditions \citep{diaconis1979}. In contrast, the conjugate family for an alternative parametrization as defined in \eqref{pifam} can be empty; see \citet[][Example 1.2]{GS95}. In what follows, we revisit the Beta--Binomial setting and showcase yet another advantage of working with the canonical parametrization.

\begin{example}\label{bbrevisited}\normalfont 
  Let $\eta = \log\{\theta / (1 - \theta)\}$ be the natural parameter of $\text{Bin}(n, \theta)$ and consider the prior for $\theta$ as $\text{Beta}(a, b)$. The conjugate prior for the natural parameter is 
  \begin{equation*}
    \pi(\eta \mid a, b) = \frac{1}{B(a, b)}\exp\{a \eta - (a + b) \log(1 + exp(\eta))\}.     
  \end{equation*}
{It is readily apparent that}
\begin{equation*}
  \|\pi\| = \frac{\{B(2a, 2b)\}^{1 / 2}}{B(a, b)}, \quad a, b > 0.
\end{equation*}
More informative priors (i.e. larger values of $a$ and/or $b$) will always be more `peaked' than less informative ones, and there is no need to constrain the range of values of the hyperparameters to the set $(1/2,\infty)$, as it was the case in \eqref{normm}. Finally, note that the max-compatible prior under the canonical parametrization is $\pi(\eta \mid n_1, n - n_1)$, whereas the max-compatible prior under the parametrization used earlier in Example~\ref{bbin} was $\beta(\theta \mid 1 + n_1, 1 + n - n_1)$.
\end{example}

{There are some links between the max-compatible prior introduced in Section~\ref{maxcompatible} and {Hartigan}'s maximum likelihood prior \citep{hartigan1998}. In the context of the exponential family, {Hartigan}'s maximum likelihood prior is a uniform distribution on the canonical parameter $\eta$. Equation~\eqref{maxmax} then implies that the max-compatible prior on the canonical parameter $\pi(\eta \mid \sum_{i = 1}^nT(Y_i), n)$, 
can be regarded as a posterior derived from {Hartigan}'s maximum likelihood prior.}

\section{Extensions}\label{extensions}
\subsection{Local prior--likelihood compatibility}\label{local} 
In some cases, when assessing the level of agreement between prior and likelihood, integrating over $\Theta$ may not be feasible, but one can still assess the level of agreement over priors supported on a subset of the parameter space. {Below  $\Theta$ represents the parameter space and $\Pi$ denotes the support of the prior. More specifically,} let $\pi$ be a prior supported on $\Pi = \{\theta: \pi(\theta) > 0\} \subseteq \Theta$. We define local prior--likelihood compatibility as
\begin{equation}
  \kappa^*_{\pi, \ell} = \frac{\langle \pi, \ell \rangle^*}{\|\pi\|^* \|\ell\|^*} = 
  \frac{\langle \pi, \ell \rangle}{\|\pi\| \|\ell\|^*},
  \label{loccomp} 
\end{equation}
where $\langle \pi, \ell \rangle^* = \int_{\Pi} \pi(\theta) \ell(\theta) \, \dif \theta$, $\|\ell\|^* = \{\int_{\Pi}  \ell^2(\theta) \, \dif \theta\}^{1 / 2}$, and $\|\pi\|^* = \{\int_{\Pi}  \pi^2(\theta) \, \dif \theta\}^{1 / 2}$. {Note that
\begin{equation*}
  \langle \pi, \ell \rangle^* = \int_{\Pi} \pi(\theta) \ell(\theta) \, \dif \theta = 
  \int_{\Theta} \pi(\btheta) \ell(\btheta)\, \dif \btheta = 
  \langle \pi, \ell \rangle,
\end{equation*}
and thus if $\Pi = \Theta$, then $\kappa^*_{\pi, \ell} = \kappa_{\pi, \ell}$.} In practice, we recommend using standard likelihood--prior compatibility \eqref{kappa.pi} instead of its local version \eqref{loccomp}, with the exception of situations for which the likelihood is square integrable over $\Pi$ but not over $\Theta$. To illustrate that \eqref{loccomp} could be well defined even if \eqref{kappa.pi} is not, suppose $Y \mid \mu, \sigma^2 \sim \text{N}(\mu, \sigma^2)$ with $\mu \sim \text{N}(m, s^2)$ and $\sigma \sim \text{Unif}(a, b)$, for $0 < a < b$. In this pathological single-observation case \eqref{kappa.pi} would not be defined, while it follows that,
\begin{equation*}
  \kappa^*_{\pi, \ell} = 
  \frac{\int_{a}^b \int_{-\infty}^{\infty} \phi(\mu \mid m, {s^2})/(b - a) \ell(\mu, \sigma) \, \dif \mu \, \dif \sigma}{[\log(b/a)/\left\{4 \pi s (b-a)\right\} ]^{1/2}}.
\end{equation*}
Since \eqref{kappa.pi} only assesses the level of agreement locally---that is, over $\Pi \subseteq \Theta$---the values of \eqref{kappa.pi} and \eqref{loccomp} are not directly comparable. A local $\kappa_{\ell, p}^*$ can be analogously defined to \eqref{loccomp}. 

\subsection{Affine-compatibility}\label{ac}
We now comment on a version of our geometric setup where one no longer focuses directly on angles between priors, likelihoods, and posteriors, but on  {functions} of these.
Specifically, we consider the following measures of agreement,
\begin{equation}\label{aff}
  \begin{cases}\displaystyle
    \kappa_{\sqrt{\pi}, \sqrt{\ell}} = \frac{\langle \sqrt{\pi}, \sqrt{\ell} \rangle}{\|\sqrt{\ell}\|}, \quad 
    \kappa_{\sqrt{\pi}, \sqrt{p}} = \langle \sqrt{\pi}, \sqrt{p} \rangle, \\ 
    \kappa_{\sqrt{\pi_1}, \sqrt{\pi_2}} = \langle \sqrt{\pi_1}, \sqrt{\pi_2} \rangle, \quad 
    \kappa_{\sqrt{p_1}, \sqrt{p_2}} = \langle \sqrt{p_1}, \sqrt{p_2} \rangle.
  \end{cases}
\end{equation}
Some affine-compatibilities in \eqref{aff} are Hellinger affinities \citep[][p.~211]{V98}, and thus have links with \cite{kurtek2015} and \cite{roos2015}.  Action does not always takes place at the Hilbert sphere, given the need of considering $\kappa_{\sqrt{\pi}, \sqrt{\ell}}$. Local versions of prior--likelihood and likelihood--posterior affine-compatibility, $\kappa_{\sqrt{\pi}, \sqrt{\ell}}$ and $\kappa_{\sqrt{\ell}, \sqrt{p}}$, can be readily defined using the same principles as in Section~\ref{local}.


It is a routine exercise to prove that max-compatible hyperparameters also maximise $\kappa_{\sqrt{\pi}, \sqrt{\ell}}$, and thus all comments on Section~\ref{maxcompatible} also apply to prior--likelihood affine-compatibility. {In terms of affine-compatibility in the exponential family, following the same notation as in Section~\ref{expf}, it can be shown that 
\begin{equation}
  \begin{cases}\displaystyle
    \kappa_{\sqrt{\pi}, \sqrt{\ell}}(\tau, n_0) = \frac{\{K(\tau, n_0) K(\sum_{i = 1}^n T(Y_i), n)\}^{1/2}}{K(1/2\{\tau + \sum_{i = 1}^n T(Y_i)\}, \{n_0 + n\}/2)}, \\ \displaystyle
    \kappa_{\sqrt{\pi}, \sqrt{p}}(\tau, n_0) =  \frac{\{K(\tau, n_0) K(\tau + \sum_{i = 1}^n T(Y_i), n_0 + n)\}^{1/2}}{K(\tau + 1/2\sum_{i = 1}^n T(Y_i), n_0 + n/2)}, \\ \displaystyle
    \kappa_{\sqrt{p}, \sqrt{\ell}}(\tau, n_0) = \frac{\{K(\tau + \sum_{i = 1}^n T(Y_i), n_0 + n) K(\sum_{i = 1}^n T(Y_i), n)\}^{1/2}}{K(1 / 2\tau + \sum_{i = 1}^n T(Y_i), n_0/2 + n)},    
  \end{cases}
\label{kapfama}
\end{equation}
with $K(\tau, n_0)$ as defined in \eqref{K}.}

Affine-compatibility between priors and posteriors is invariant to monotone increasing parameter transformations, as a consequence of properties of the Hellinger distance \cite[][p.~267]{roos2011}.
Affine-compatibility counterparts of all data examples are available from the supplementary materials; the conclusions are tantamount to the ones using compatibility.

\section{Posterior and prior mean-based estimators of compatibility}\label{estimation}
In many situations closed form estimators of $\kappa$ and $\| \cdot \|$ are not available. This leads to considering algorithmic techniques to obtain estimates. As most Bayes methods resort to MCMC methods it would be appealing to express $\kappa_{\cdot,\cdot}$ and $\| \cdot \|$ as functions of posterior expectations and employ MCMC iterates to estimate them. For example, $\kappa_{\pi,p}$ can be expressed as   
\begin{equation}\label{EstKappaE1}
    \kappa_{\pi,p} = E_p \, \pi(\btheta)
     \left[E_p \left\{\frac{\pi(\btheta)}{\ell(\btheta)} 
      \right\} E_p \{\ell(\btheta)\pi({\theta})\} \right]^{-1/2},
\end{equation}
where $E_p(\,\cdot\,)=\int_{\Pi} \cdot\, p(\btheta \mid {y}) \, \dif \btheta$ is the expected value with respect to the posterior density. A natural Monte Carlo estimator would then be 
\begin{equation}\label{hme}
  \begin{split}
    \hat{\kappa}_{\pi,p} = \frac{1}{B}\sum_{b=1}^B  \pi({\theta}^b) 
    \bigg[  \bigg\{\frac{1}{B}\sum_{b=1}^B\frac{\pi({\theta}^b)}{\ell({\theta}^b)}\bigg\}   \bigg\{\frac{1}{B}\sum_{b=1}^B\ell({\theta}^b)\pi({\theta}^b)\bigg\}\bigg]^{-1/2},
   \end{split}
\end{equation}
where ${\theta}^b$ denotes the $b$th MCMC iterate of $p({\theta} \mid {y})$. Consistency of such an estimator follows trivially by the ergodic theorem and the continuous mapping theorem, but there is an important issue regarding its stability. Unfortunately, \eqref{EstKappaE1} includes an expectation that contains $\ell({\theta})$ in the denominator and therefore \eqref{hme} inherits the undesirable properties of the so-called harmonic mean estimator \citep{Newt:Raft:appr:1994}. It has been shown that even for simple models this estimator may have infinite variance (\citealt{RafteryNewtonSatagopanKrivitsky:2007}), and has been harshly criticized for, among other things, converging extremely slowly. Indeed, as argued by \citet[][p.~655]{wolpert2012alpha}: 
    ``the reduction of Monte Carlo sampling error by a factor of two requires increasing the Monte Carlo sample size by a factor of $2^{1/\varepsilon}$, or in excess of $2.5\cdot 10^{30}$ when $\varepsilon=0.01$, rendering [the harmonic mean estimator] entirely untenable.''  


\begin{figure}[h]
  \centering 
  \includegraphics[height = 6.3cm]{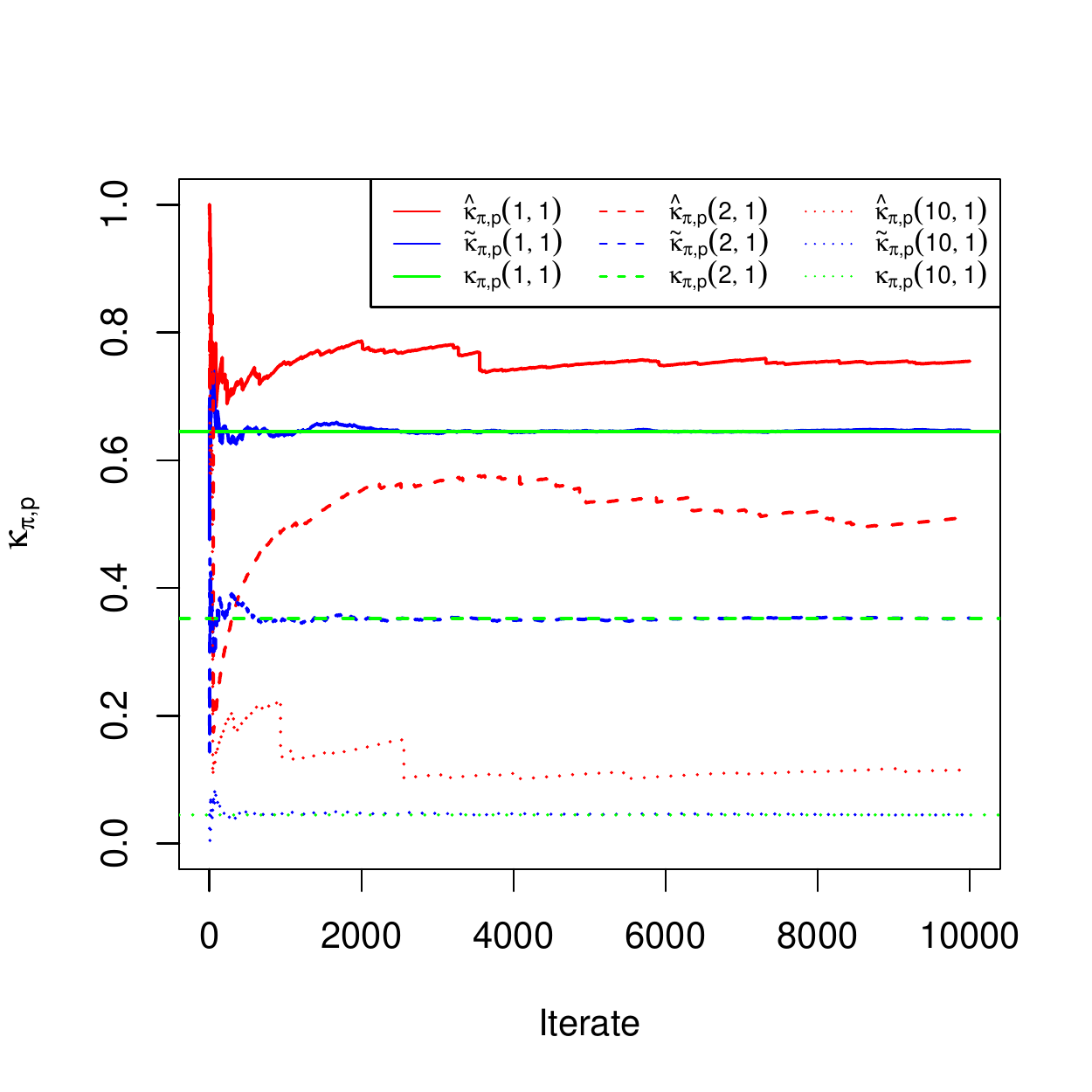}
  \caption{\footnotesize Running point estimates of prior--posterior compatibility, $\kappa_{\pi,p}$, for the {on-the-job drug usage toy example}. Green lines correspond to the true $\kappa_{\pi, p}$ values computed as in Example~\ref{exkpip}, blue represents $\tilde{\kappa}_{\pi,p}$ and red denotes  $\hat{\kappa}_{\pi,p}$. Notice that $\tilde{\kappa}_{\pi,p}$ converges to the true $\kappa_{\pi,p}$ values quickly while $\hat{\kappa}_{\pi,p}$ will need much more than 10\,000 Monte Carlo draws to converge.}
\label{KhKt}
\end{figure}

An alternate strategy is to avoid writing $\kappa_{\pi,p}$ as a function of harmonic mean estimators and instead express it as a function of posterior and prior expectations. For example, consider
\begin{equation}\label{kappa.E2}
    \kappa_{\pi,p} = 
    E_p \, \pi(\btheta)
    \left[\frac{E_{\pi}\{\pi(\btheta)\}}{E_{\pi}\{\ell({\theta})\}} \, E_p\{\ell({\theta})\pi({\theta})\} \right]^{-1/2},
\end{equation}
where $E_{\pi}(\,\cdot\,)=\int_{\Pi} \cdot\, \pi(\btheta) \, \dif \btheta$. Now the Monte Carlo estimator is  
\begin{equation}\label{hme2}
  \begin{split}
    \tilde{\kappa}_{\pi,p} = \frac{1}{B}\sum_{b=1}^B \pi({\theta}^b)  \bigg[\bigg\{\frac{\sum_{b=1}^B\pi({\theta}_b)}{\sum_{b=1}^B\ell({\theta}_b)}\bigg\}   \bigg\{\frac{1}{B}\sum_{b=1}^B\ell({\theta}^b)\pi({\theta}^b)\bigg\}\bigg]^{-1/2},
  \end{split}
\end{equation}
where ${\theta}_b$ denotes the $b$th draw of ${\theta}$ from $\pi({\theta})$, which can also be sampled within the MCMC algorithm. {Although representations \eqref{kappa.E2} and \eqref{hme2} could in principle suffer from numerical instability for diffuse priors, they behave much better in practice than \eqref{EstKappaE1} and \eqref{hme}.}  To see this, Figure~\ref{KhKt} contains running estimates of $\kappa_{\pi,p}$ using \eqref{hme} and \eqref{hme2} for Example \ref{ex2} with three prior parameter specifications, namely: $(a=1, b=1)$, $(a=2, b=1)$, and $(a=10, b=1)$; the true $\kappa_{\pi, p}$ for each prior specification is also provided. It is fairly clear that $\hat{\kappa}_{\pi,p}$ displays slow convergence and large variance, while $\tilde{\kappa}_{\pi,p}$ converges quickly. 

The next proposition contains prior and posterior mean-based representations of geometric quantities that can be readily used for constructing Monte Carlo estimators.  

\begin{proposition}\label{postmean}
  Let $\pi$ be a prior supported on $\Pi = \{\theta: \pi(\theta) > 0\} \subseteq \Theta$, with  $\|\ell\|^*$ and $\kappa_{\pi,\ell}^*$ be defined as in \eqref{loccomp}, and let $E_p(\,\cdot\,)=\int_{\Pi} \cdot\, p(\btheta \mid {y})
    \, \dif \btheta$ and $E_{\pi}(\,\cdot\,) = \int_{\Pi} \cdot \ \pi(\btheta) \, \dif \btheta$. Then,
    \begin{equation*}
      \begin{split}
        \|p\| & = \bigg\{\frac{E_{p}\{\ell(\btheta)
          \pi({\theta})\}}{E_{\pi} \, \ell(\btheta)}\bigg\}^{1 / 2}, \quad
        \|\pi\| = \{E_{\pi} \, \pi({\theta})\}^{1 / 2}, \quad 
        \|\ell\|^* = \bigg\{E_{\pi}\, \ell({\theta}) \,
        E_p\left\{\frac{\ell({\theta})}{\pi({\theta})}\right\}\bigg\}^{1 / 2},\\[0.2cm]\nonumber
        \kappa_{\pi,\ell}^*  &= E_{\pi} \, \ell(\btheta) \bigg[E_{\pi} \, \pi({\theta}) \, E_{\pi} \, \ell(\btheta) \, E_p\left\{\frac{\ell(\btheta)}{\pi({\theta})}\right\}\bigg]^{-1/2}, \quad
        \kappa_{\pi,p}  = E_{p}\, \pi({\theta})
        \bigg[\frac{E_{\pi} \, \pi({\theta})}{E_{\pi}\,
          \ell(\btheta)}E_p\left\{\ell(\btheta)\pi({\theta})\right\}\bigg]^{-1/2}, \\
        \kappa_{\pi_1,\pi_2} & = E_{\pi_1}\, \pi_2({\theta})
        \bigg[E_{\pi_1}\, \pi_1({\theta}) \, E_{\pi_2}\,
        \pi_2({\theta}) \bigg]^{-1/2}, \quad 
 \kappa_{\ell, p}^* = E_p \, \ell(\theta)  \bigg[E_{p}\left\{\frac{\ell(\btheta)}{\pi({\theta})}\right\}E_p\left\{\ell(\btheta)\pi({\theta})\right\}\bigg]^{-1/2}.
      \end{split}
\end{equation*}

\end{proposition}
Similar derivations can be used to obtain posterior and prior mean-based estimators for affine-compatibility; see supplementary materials. In the next section we provide an example that requires the use of Proposition~\ref{postmean} to estimate $\kappa$ and $\| \cdot \|$.

\section{Example: Regression shrinkage priors}\label{Reg}
\subsection{Compatibility of Gaussian and Laplace priors}
The linear regression model is ubiquitous in applied statistics. In vector form, the model is commonly written as 
\begin{equation}\label{regressionModel}
  {y} = {X} {\beta} + {\varepsilon}, \quad {\varepsilon} \sim \text{N}({0}, \sigma^2{I}),
\end{equation}
{where ${y} = (Y_1, \dots, Y_n)^{\T}$}, ${X}$ is a $n\times p$ design matrix, ${\beta}$ is a $p$-vector of regression coefficients, and $\sigma^2$ is an unknown idiosyncratic variance parameter{; the experiments below employ $\sigma\sim \text{Unif}(0, 2)$.} We consider Gaussian and Laplace prior distributions for ${\beta}$.  As documented in \cite{Park:2008} and \cite{Kyung:2010} ridge regression and $\beta_j \iid \text{N}(0, \lambda^2)$  produce the same regularization on ${\beta}$ while the lasso  produces the same regularization on $\bm{\beta}$ as assuming $\beta_j \iid \text{Laplace}(0, b)$ (where $\var(\beta_j) = 2b^2$). Below, we use $\pi_1$ to denote a Gaussian prior and $\pi_2$ a Laplace. Further,  we set $b = \sqrt{0.5\lambda^2}$ which ensures that  $\var_{\pi_1}(\beta_j) = \var_{\pi_2}(\beta_j) = \lambda^2$ for all $j$.

\subsection{Prostate cancer data example} \label{prostrate.cancer.example}
We now consider the prostate cancer data example found in \citet[][Section~3.4]{HTF:2008} to explore the `informativeness' of and various compatibility measures  for $\pi_1$ and $\pi_2$. In this example the response variable is the level of prostate-specific antigens measured on 97 males. Eight other clinical measurements (such as age and log prostate weight) were also measured and are used as covariates.  

\begin{figure}[h]
  \centering
  \begin{minipage}[c]{6cm}\centering
    \includegraphics[height=6.3cm]{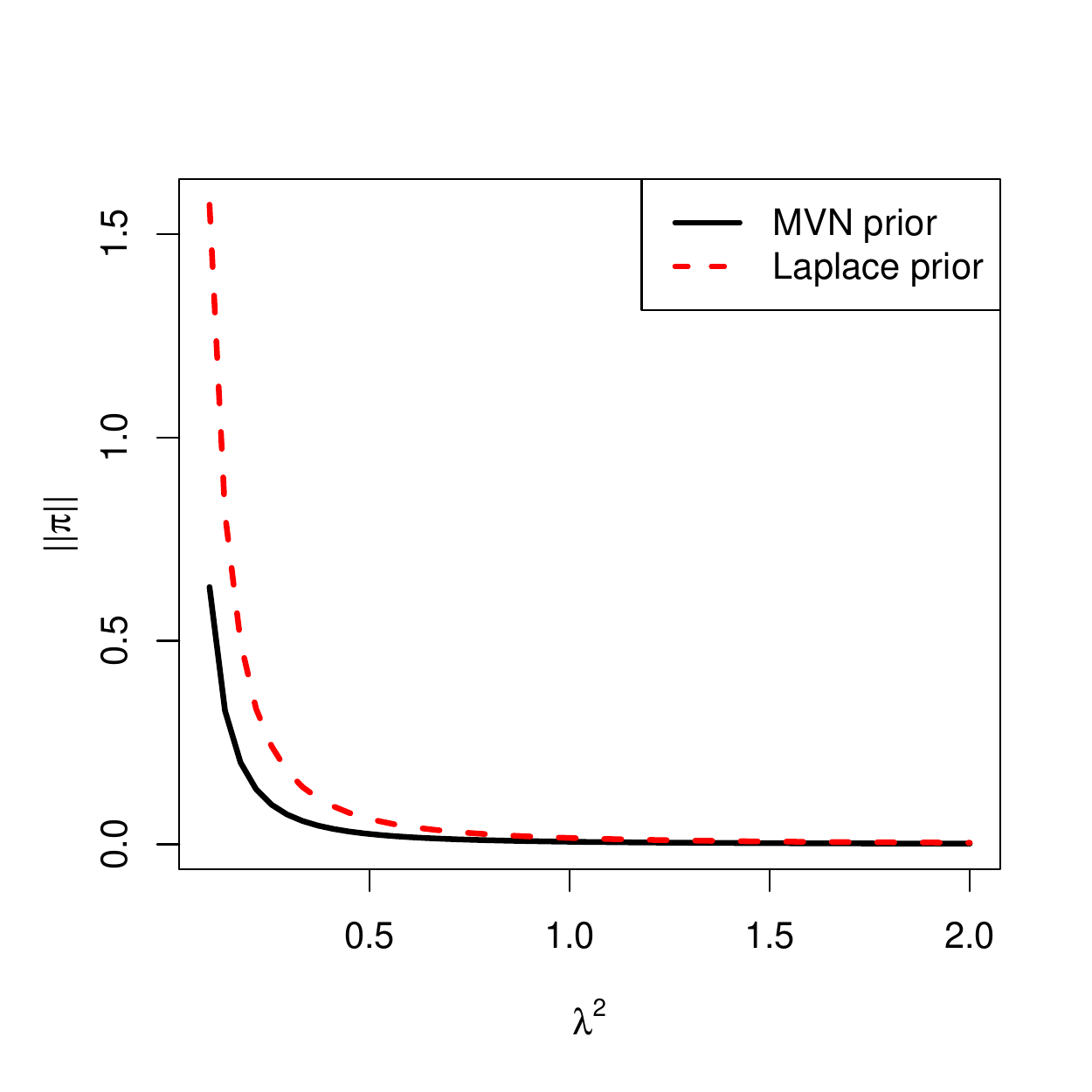} 
  \end{minipage}\hspace{1cm}
  \begin{minipage}[c]{6cm}\centering
    \includegraphics[height=6.3cm]{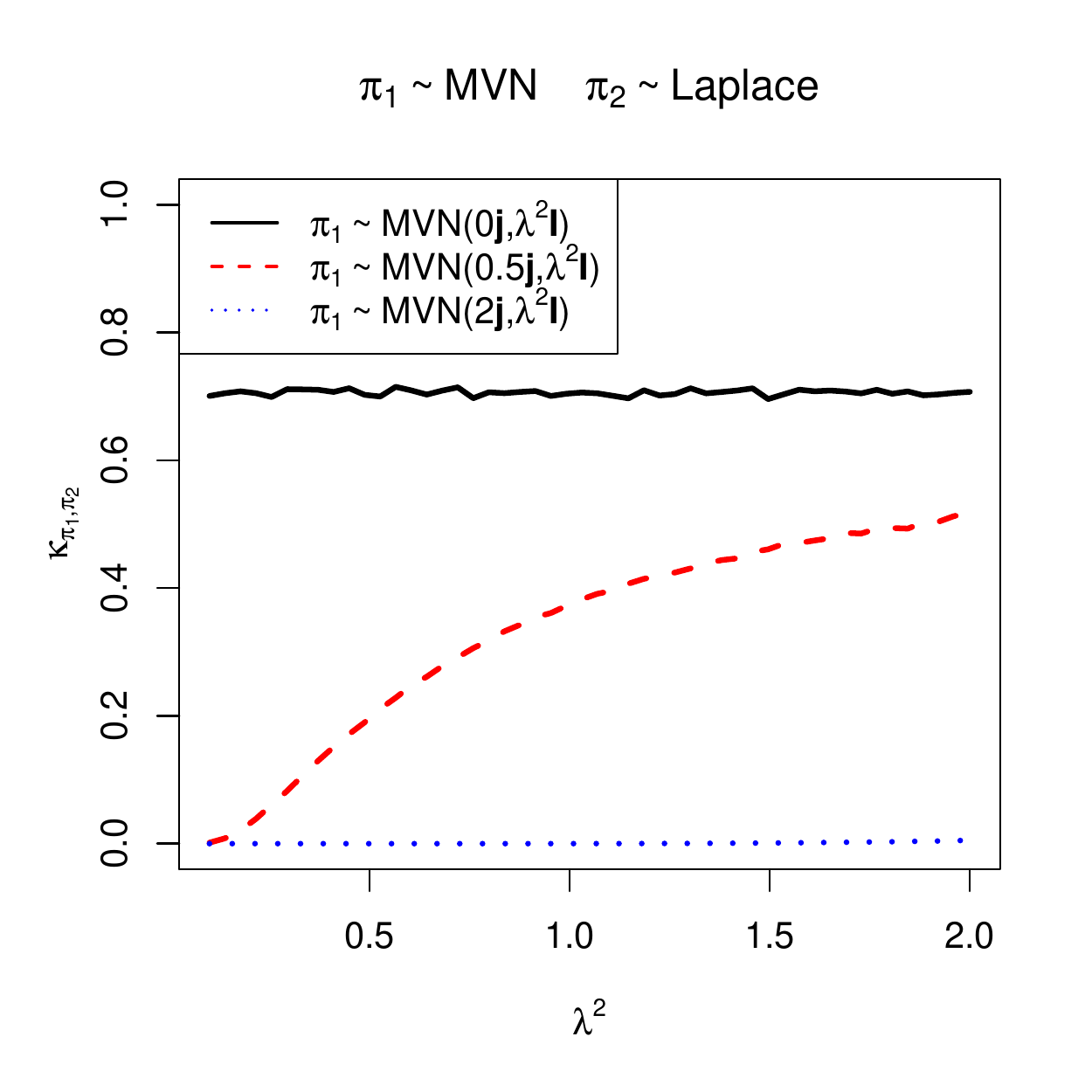} 
  \end{minipage} 
\caption{\footnotesize A comparison of priors associated with Ridge (MVN, $\pi_1$) and Lasso (Laplace, $\pi_2$) regularization in regression models in terms of $\|\pi\|$ and $\kappa_{\pi_1, \pi_2}$. The left plot depicts $\| \cdot \|$ as a function of $\lambda^2$ for both $\pi_1$ and $\pi_2$. The right compares $\kappa_{\pi_1, \pi_2}$ values as a function of $\lambda^2$ when $\pi_1$ and $\pi_2$ are centered at zero to that when the center of $\pi_1$ moves away from zero.}
\label{MVNLapNormPrioKap}
\end{figure}

We first evaluate the `informativeness' of the two priors  by computing $\| \pi_1 \|$ and $\| \pi_2 \|$ and then their compatibility using $\kappa_{\pi_1,\pi_2}$. All calculations employed Proposition~\ref{postmean} and results for a sequence of $\lambda^2$ values are provided in Figure~\ref{MVNLapNormPrioKap}.  Focusing on the left plot of Figure~\ref{MVNLapNormPrioKap} it appears that for small values of the $\lambda^2$, $\| \pi_1 \| < \| \pi_2\|$, indicating that the Laplace prior is more peaked than the Gaussian. Thus, even though the  Laplace has thicker tails, it is more `informative' relative to the Gaussian. This corroborates the lasso penalization's ability to shrink coefficients to zero (something ridge regulation lacks). As $\lambda^2$ increases the two norms converge as both spread their mass more uniformly. The right plot of Figure~\ref{MVNLapNormPrioKap}  depicts $\kappa_{\pi_1, \pi_2}$ as a function of $\lambda^2$. When $\pi_1$ is centered at zero, then  $\kappa_{\pi_1, \pi_2}$ is constant over values of $\lambda^2$ which means that mass intersection when both priors are centered at zero is not influenced by tail thickness. Compare this to $\kappa$ values  when $\pi_1$ is not centered at zero [i.e., $\pi_1 \sim \text{MVN}(0.5{j}, \lambda^2{I})$ or $\pi_1 \sim \text{MVN}(2{j}, \lambda^2{I})$]. For the former, $\kappa$ increases as intersection of prior and posterior mass increases. For the latter, $\lambda^2$ must be greater than two for there to be any substantial mass intersection as $\kappa_{\pi_1, \pi_2}$ remains essentially at zero.

\begin{figure}
\begin{center}
\includegraphics[height=6.3cm]{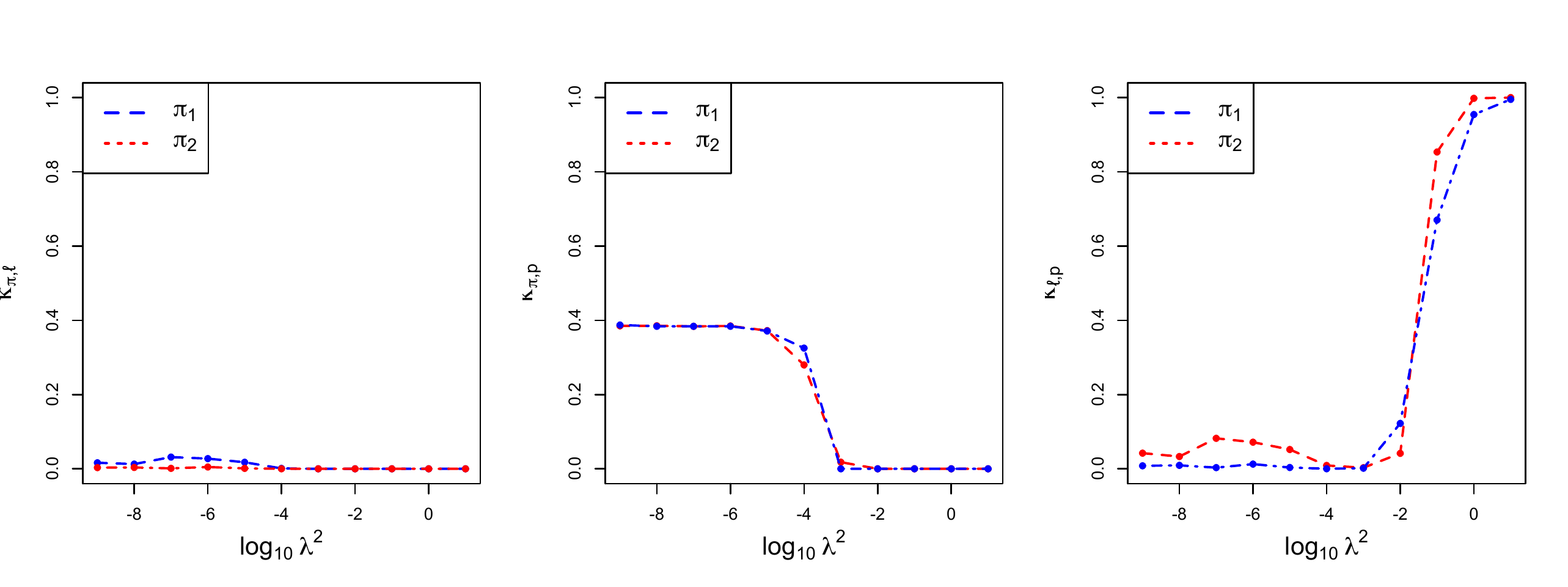}
\caption{\footnotesize Compatibility ($\kappa$) for linear regression model in \eqref{regressionModel}, with shrinkage priors, applied to the prostrate cancer data from \citet[][Section~3.4]{HTF:2008}. The $\kappa$ estimates were computed using Proposition~\ref{postmean}.}
\label{RegressionKappa}
\end{center}
\end{figure}

We now fit model \eqref{regressionModel} to the cancer data and use Proposition~\ref{postmean} to calculate various measures of compatibility. Without loss of generality we centered the ${y}$ so that ${\beta}$ does not include an intercept and standardized each of the eight covariates to have mean zero and standard deviation one. {The results are available from Figure~\ref{RegressionKappa}.}

Focusing on the left plot of Figure~\ref{RegressionKappa} the small values of $\kappa_{\pi_1, \ell}$ and  $\kappa_{\pi_2, \ell}$ indicate the existence of prior--data incompatibility. For small values of $\lambda^2$,  $\kappa_{\pi_1, \ell} > \kappa_{\pi_2, \ell}$ indicating  more compatibility between prior and data for the Gaussian prior.  {Prior--posterior compatibility} ($\kappa_{\pi, p}$) is very similar for both priors with that for $\pi_2$ being slightly smaller when $\lambda^2$ is close to $10^{-4}$.  The slightly higher $\kappa_{\pi, p}$ value for the Gaussian prior implies that it has slightly more influence on the posterior than the Laplace. Similarly, the Laplace prior seems to produce larger $\kappa_{\ell, p}$ values than that of the Gaussian prior and $\kappa_{ \ell,p_2}$ approaches one quicker than $\kappa_{\ell, p_1}$ {indicating a larger amount of posterior-data compatibility}.     Overall, it appears that the Gaussian prior has more influence on the resulting posterior distribution relative to the Laplace when updating knowledge via Bayes theorem. Similar conclusions as above would be reached by considering affine-compatibility; see supplementary materials.

\section{Discussion}\label{discussion}
Bayesian inference is regarded from the viewpoint of the geometry of Hilbert spaces. The framework offers a direct connection to Bayes theorem, and a unified treatment that can be used to quantify the level of agreement between priors, likelihoods, and posteriors---or {functions} of these. The possibility of developing new probabilistic models, obeying the geometrical principles discussed here, offering alternative ways to recast the prior vector using the likelihood vector remains to be explored. In terms of high-dimensional extensions, one could anticipate that as the dimensionality increases, there is increased potential for disagreement between two distributions. Consequently, $\kappa$ would generally diminish as additional parameters are added, \textit{ceteris paribus}, but a suitable offsetting transformation of $\kappa$ could result in a measure of `per parameter' agreement. 

{Some final comments on related constructions are in order. Compatibility as set in Definition~\ref{Comp.def} includes as a particular case the measures of niche overlap in \cite{ss1980}. Peakedness as discussed in here should not be confused with the concept of \cite{B48}. The geometry in Definition~\ref{absgeom} has links with the so-called affine space and thus the geometrical framework discussed above is different but has many similarities with that of \cite{marriott2002} and also with the mixture geometry of \cite{amari2016}. A key difference is that the latter approaches define an inner product with respect to a density which is the basis of the construction of the Fisher information while here we define it simply as the product of two functions in $L_2(\Theta)$, and connect the construction with Bayes theorem and with Pearson's correlation coefficient. While here we deliberately focus on positive $g, h \in L_2(\Theta)$, the case of a positive $m \equiv g(\theta) + k h(\theta) \in L_2(\Theta)$---but with $g$ always positive and with $h$ negative on a part of $\Theta$---is of interest in itself, as well as the set values of $k$ ensuring positivity of $m$ for all $\theta$. Some further interesting setups would be naturally allowed by slightly extending our geometry, say to include `mixtures' with negative weights. Indeed, the parameter $\lambda$ in \eqref{lambda} might in some cases be allowed to take some negative values while the resultant function is still positive; see \cite{anaya2007}.}

{While not explored here, the use of compatibility as a means of assessing the suitability of a given sampling model, is a natural inquiry for  future research.}

\vspace{0.2cm} \noindent \textbf{Supplementary material}: 
The online supplementary materials include the counterparts of the data examples in the paper for the case of affine-compatibility as introduced in Section~\ref{ac}, technical derivations, and proofs of propositions.

\renewcommand\refname

\end{document}